\documentclass[a4paper,11pt]{article}
\pdfoutput=1 

\usepackage{jcappub} 
\usepackage[T1]{fontenc} 
\usepackage[utf8]{inputenc}

\usepackage{hyperref} 
\usepackage{url}
\usepackage{graphicx}
\usepackage{esvect}
\usepackage{graphics}
\usepackage{natbib}
\usepackage{mathrsfs}
\usepackage{blindtext}

\usepackage{lineno}

\usepackage[table]{xcolor}
\usepackage{multirow}
\usepackage{placeins}

\usepackage[normalem]{ulem}
\usepackage{color}
\usepackage{soul}

\usepackage{supertabular}
\usepackage{tabularx}
\usepackage{comment}

\def\units#1{~\hbox{$\,{\rm #1}$}}

\title{Markov chain Monte Carlo analyses of the flux ratios of B, Be and Li with the DRAGON2 code
}

\author[a,b]{P.~De~La~Torre~Luque}
\emailAdd{pedro.delatorreluque@ba.infn.it}
\author[a]{M.~N.~Mazziotta}
\emailAdd{mazziotta@ba.infn.it}
\author[a,b]{F.~Loparco}
\emailAdd{Francesco.Loparco@ba.infn.it}
\author[a]{F.~Gargano}
\emailAdd{Fabio.Gargano@ba.infn.it}
\author[a,b]{D.~Serini}
\emailAdd{Davide.Serini@ba.infn.it}

\affiliation[a]{Istituto Nazionale di Fisica Nucleare, Sezione di Bari, via Orabona 4, I-70126 Bari, Italy}
\affiliation[b]{Dipartimento di Fisica ``M. Merlin" dell'Universit\`a e del Politecnico di Bari, via Amendola 173, I-70126 Bari, Italy}

\date{\today}

\abstract{Recent cosmic-ray measurements are challenging our models of propagation in the Galaxy. A good characterization of the secondary cosmic rays (B, Be, Li and sub-iron species) is crucial to constrain these models and exploit the precision of modern CR experiments. In this work, a Markov chain Monte Carlo analysis has been implemented to fit the experimental flux ratios between B, Be and Li and their flux ratios to the primary cosmic-ray nuclei C and O. We have fitted the data using two different parametrizations for the spallation cross sections. The uncertainties in the evaluation of the spectra of these secondary cosmic rays, due to spallation cross sections, have been taken into account by introducing scale factors as  nuisance parameters in the fits, assuming that this uncertainty is mostly due to the normalization of the cross sections parametrizations. We have also tested two different formulations for the diffusion coefficient, which differ in the origin of the high energy hardening ($\sim 200\units{GeV/n}$) of cosmic rays. Additionally, two different approaches are used to scale the cross sections, one based on a combined analysis of all the species (``combined'' analysis) and the other reproducing the high energy spectra of the secondary-to-secondary flux ratios of Be/B, Li/B, Li/Be (``scaled'' analysis). This allows us to make a better comparison between the propagation parameters inferred from the cross sections parametrizations tested in this work. This novel analysis has been successfully implemented using the numerical code DRAGON2 dedicated to cosmic-ray propagation to reproduce the cosmic-ray nuclei data up to $Z=14$ from the AMS-02 experiment. 
In general, it is found that the ratios of Li favor a harder spectral index of the diffusion coefficient, but compatible with the other ratios inside the observed $2\sigma$ uncertainties. In addition, it is shown that, including these scale factors, the secondary-to-primary flux ratios can be simultaneously reproduced, obtaining that the scale factor associated to the cross sections of boron production is the lowest one, whereas that associated to Li production is the largest one.
}

\begin{document}
\maketitle
\flushbottom

\section{Introduction}
\label{sec:intro}

The high precision measurements provided by recent cosmic-ray missions such AMS-02~\cite{Aguilar:2015ooa, Aguilar:2014mma, Aguilar:2016kjl} and  DAMPE~\cite{Ambrosi:2017wek, DAMPE_protons} together with the increasing quality and quantity of gamma-ray data, thanks to the Fermi-LAT~\cite{Ackermann_2012, Ackermann:2014usa}, HAWC~\cite{Albert:2020fua, Abeysekara:2018bfb, Malone:2020vcs} and H.E.S.S.~\cite{Aharonian_2008, Egberts:2011zz, Aharonian_2006}, are requiring improved models on particle acceleration and transport for their interpretation~\cite{Serpico:2018lkb}.
The transport of Galactic cosmic rays (CRs) is governed by their interactions with the magneto-hydrodynamical turbulence generated in the plasma of the interstellar medium (ISM). This mechanism is responsible for the observed isotropy in their arrival directions at Earth~\cite{Ginz&Syr,berezinskii1990astrophysics} and is usually described with a diffusive transport equation.

Injection and acceleration of primary CRs is believed to take place mainly in supernova remnants (SNRs)~\cite{blasi2013origin} (although also pulsars~\cite{Hooper:2008kg, Kotera:2015rga} and OB stars~\cite{Binns:2007yu} are likely CR sources)  and it is explained, at least in general principles, by the diffusive shock acceleration (DSA) model~\cite{krymskii1977regular,bell1978acceleration, axford1982structure, blandford1978particle}. 
In turn, the study of secondary CRs (i.e. those generated from the spallation reactions of primary CRs with gas nuclei in the ISM) provides 
crucial information about the propagation of CRs; the amount of matter traversed (grammage) or path length~\cite{Blasi:2017caw, stephens1998cosmic}, the size of the magnetised halo~\cite{bringmann2012radio, biswas2018constraining, moskalenko2000diffuse} or the spectrum of self-generated turbulence~\cite{MORLINO201839, Aloisio:2013tda} can, in fact, be inferred from the measured fluxes of secondary CR nuclei.

In particular, the flux ratio of secondary to primary CRs is sensitive to the propagation parameters~\cite{reinert2018precision, derome2019fitting}, whereas it is mostly insensitive to the source spectrum of primary species. As a first order approximation, the flux ratio between the $k$-th species of secondary CRs and the $i$-th species of primaries can be written as:
\begin{equation}
\frac{J_{k}}{J_{i}}\left(R\right) \approx \frac{\tau_{diff}(E) Q_{k}(E)}{J_{i}(R)}\sim \sigma_{i \rightarrow k}(R) \tau_{diff}(R) \propto \sigma_{i \rightarrow k}(R)/D(R)
\label{eq:sec/prim}
\end{equation}
where $R$ is the rigidity of CR particles.
In the previous equation we indicate with $Q_{k} = n_{ISM} \sigma_{i \rightarrow k} J_{i}$ the source term for secondary CRs, where $\sigma_{i \rightarrow k}$ is the production cross section of the secondary CR species $k$ from the interaction of the primary CR species $i$ with the gas in the ISM (spallation reactions) and $n_{ISM}$ is the density of gas in the ISM (consisting of H:He = 1:0.1 and traces of heavier elements~\cite{Composition}). Finally, $\tau_{diff} \propto 1/D(R)$ is the mean diffusion time of CRs in the Galaxy, where $D(R)$ is the diffusion coefficient used to describe CR propagation in the Galaxy.

The spallation cross sections are one of the main ingredients in CR propagation computations, since they are crucial for evaluating the fluxes of different secondary CR species. However, due to the lack of experimental data and to the large number of interaction channels involved in the production of each isotope, the uncertainties on the cross sections are still quite large. These uncertainties propagate to the predicted fluxes of secondary CRs, which are therefore also uncertain.
In order to reduce the impact of the lack of data, we make use of parametrizations and extrapolations of the spallation cross sections involved in the CR network~\cite{Tomassetti:2017hbe}. This can be done due to the fact that the energy dependence of the cross sections is generally better constrained and most of the uncertainties in the parametrizations are mainly dominated by their normalization, as recently quantified in ref.~\cite{CarmeloBlasi}.

Recently, some authors have proposed to use other secondary-to-primary flux ratios to cross check the validity of the cross sections parametrizations~\cite{maurin2010systematic}, concluding that this cross check will help reducing the systematic uncertainties related to the determination of the propagation parameters. In this work, we have implemented two different and well known cross sections parametrizations in a customised version\footnote{This version is public at \url{https://github.com/tospines/Customised-DRAGON2_beta/}}~\cite{pedro_de_la_torre_luque_2021_4461732} of the {\tt DRAGON2} CR propagation code \cite{DRAGON2-1, DRAGON2-2}.
We have then optimized the propagation parameters to reproduce the AMS-02 experimental data on secondary-to-primary flux ratios of B, Be and Li to C and O as well as the corresponding secondary-to-secondary flux ratios. 
In this way we study how spallation cross sections parametrizations affect the determination of the propagation parameters for the different secondary CRs. This work is the follow-up of our recent paper~\cite{Luque:2021joz}, where we investigated how the choice of cross sections affects the predictions on the fluxes of the light secondary CRs B, Be and Li, and demonstrated that a simultaneous fit of the fluxes of these secondary CRs can be achieved above a few GeV/n taking into account the uncertainties in the normalization of the cross sections used.

There are very few works involving the ratios of secondary Li and Be nuclei (see~\cite{Weinrich:2020cmw, DENOLFO20061558}). This is mainly due to the fact that experimental fluxes of Li and Be were poorly measured before AMS-02 (for a comparison, see ref.~\cite{AMS_LiBeB}). In addition, the uncertainties on the predicted fluxes of Li and Be~\cite{DENOLFO20061558} are larger than those on the flux of B. In this work, two different strategies have been adopted to reduce the systematic uncertainties related to the the spallation cross sections parametrizations using combined analyses of the secondary-to-primary flux ratios with a nuisance parameter to account for the cross sections normalization uncertainties and the secondary-to-secondary flux ratios between B, Be and Li.

This paper is organised as follows: in section~\ref{sec:bayes} we illustrate the details of the Markov chain Monte Carlo (MCMC) algorithm and the analyses performed for the flux ratios of B, Be and Li to C and O (B/C, B/O, Be/C, Be/O, Li/C and Li/O). Then, in section~\ref{sec:results}, we report and discuss the results on the determination of the diffusion parameters using what we call ``standard analyses'' for the different diffusion coefficient and cross sections parametrizations adopted, taking into account the uncertainties related to the normalization of spallation cross sections for Be, B and Li production by means of the use of nuisance parameters. In section~\ref{sec:CXSanal}, we employ another approach that allows us to compare the diffusion parameters found for different secondary CRs after tuning the normalization of the cross sections parametrizations in a similar way as explored in~\cite{Luque:2021joz}. Finally, in sect.~\ref{MCMC_conc} we discuss the main conclusions of this work.

\section{Determination of the propagation parameters}
\label{sec:bayes}

A spectral break (the rigidity where the CR spectrum changes its slope) at around $300 \units{GV}$, has been observed by many experiments in primary and secondary species~\cite{Vladimirov:2011rn, Aguilar:2015ooa, aguilar2015precision, Aguilar:2018njt, eaax3793}. Although a quantitative analysis of B/C data favours a scenario with a break due to a diffusion mechanism (``diffusion hypothesis'')~\cite{genolini2017indications, Weinrich:2020cmw}, the scenario of a break in the injection spectra (``source hypothesis'') or a combination of both mechanisms (supported by the fact that different primary species point to different break positions~\cite{Niu:2020qnh}) can not be discarded. In this work we have therefore tested two well known parametrizations of the diffusion coefficient, that differ in the physical interpretation of the high energy part of the CR spectra, as explained above, and are given by the following equations: 
\begin{equation}
D = D_0 \beta^{\eta}\left(\frac{R}{R_0} \right)^{\delta} \text{ \hspace{1.8 cm} \textbf{Source hypothesis}}
\label{eq:sourcehyp}
\end{equation}
\begin{equation}
 D = D_0 \beta^{\eta}\frac{\left(R/R_0 \right)^{\delta}}{\left[1 + \left(R/R_b\right)^{\Delta \delta / s}\right]^s}
 \text{  \hspace{0.7 cm} \textbf{Diffusion hypothesis}}
\label{eq:breakhyp}
\end{equation}
where the reference rigidity is set to $R_0 = 4\units{GV}$.

As in~\cite{Luque:2021joz}, where we refer to for full details about the set-up of the CR propagation simulations, we solved the complete propagation equation using the DRAGON2 code for a diffusion-reacceleration model, adopting a 2D ISM gas distribution. In both diffusion models we use the values of the Alfvén velocity (that accounts for CR reacceleration), $V_A$, the normalization of the diffusion coefficient $D_0$, the spectral index, $\delta$, and the $\beta$ (particle's speed, in speed of light units) exponent, $\eta$, as the parameters that we will include in our analyses, and we refer to them as diffusion parameters. 
In equation~\ref{eq:breakhyp} there are three additional parameters: the rigidity break ($R_b$), the change in spectral index ($\Delta\delta = \delta - \delta_h $, where $\delta_h$ is the spectral index for rigidities above the break), and a smoothing parameter $s$, used to allow a soft transition around the hardening position. In our analyses, we fix these parameters to the values found in ref.~\cite{genolini2017indications}, which were determined from a detailed analysis of the AMS-02 fluxes of protons and helium (with less experimental uncertainties than heavier nuclei at high energies). These values are: $\Delta\delta = 0.14 \pm 0.03$, $R_b = (312 \pm 31) \units{GV}$ and $s = 0.040 \pm 0.0015$. The fact that these parameters are fixed from the primary fluxes allows us to avoid degeneracies with the evaluation of the main propagation parameters, leading to a more rigid determination. We have observed that, varying these parameters within their $1\sigma$ uncertainty values only affects the determination of the spectral index, $\delta$, by a maximum variation of $\pm 1\%$, which should be taken as an additional systematic uncertainty in the determination of $\delta$.

The mathematical model used to parametrize the injection spectra of primary nuclei is a doubly broken power law, with a high-energy break at $335\units{GeV}$ for protons and at $200\units{GeV/n}$ for heavier elements, and a low-energy break at $7\units{GeV/n}$ for all primary CRs, when we treat the diffusion parameters in the source hypothesis, whereas we use a simple broken power law when we treat the diffusion hypothesis, with low-energy break in the injection spectra at $7\units{GeV/n}$. Then, to account the solar modulation effect, we make use of the Force-field approximation~\cite{forcefield}, which is characterized by the value of the Fisk potential. We set the Fisk potential to the value found in~\cite{Luque:2021joz}, corresponding to $\phi = 0.61 \units{GV}$ (obtained from the NEWK neutron monitor experiment \footnote{\url{http://www01.nmdb.eu/station/newk/}} neutron monitor experiment - see~\cite{ghelfi2016non,ghelfi2017neutron} - in combination with Voyager-1 data~\cite{Stone150}) for the period between May, 2011 to May, 2016, that corresponds to the period of collected B, Be and Li data by the AMS-02 collaboration. The source distribution used in these computations is that from~\cite{Lorimer_2006} and the gas distribution function is taken from~\cite{ferriere2007spatial}. We argue here that, while the source distribution does not seem to have any impact on our results, the gas distribution used can have a sizeable impact on the determination of the scale factors. In particular, we observed changes in the secondary-to-secondary fluxes below 5\% at energies above $10 \units{GeV/n}$ when using different gas distributions, and these changes can be even larger at low energies 
. Furthermore, the authors of ref.~\cite{Johannesson:2018bit} report differences in the propagation parameters derived in the case of 2D and 3D spatial distributions of the gas.

Basically, our strategy consists of an iterative procedure, which starts taking a set of propagation parameters ($D_0$, $V_A$, $\eta$ and $\delta$) as first guess; then, a fit of the ratios of Li, Be and B to C and O to the AMS-02 data is performed to determine these parameters and their associated uncertainties. Before the fit, the injection spectra are adjusted by fitting the AMS-02 data for C, N, O, Ne, Mg and Si. If the guessed diffusion parameters are very different (out of 1 $\sigma$ uncertainty) from those predicted by the fit, the injection spectra are adjusted again with the new predicted set of diffusion parameters and the fit is performed again to obtain a more refined prediction on the diffusion parameters. Convergence is reached when the output diffusion parameters after the last iteration are consistent (within 1 $\sigma$ uncertainty) with those used in the previous iteration to adjust the injection spectra. As expected, one-two iterations are needed, since the secondary-to-primary flux ratios are highly independent of the injection parameters. In this way, the injection spectra of primary CRs are always adjusted to reproduce AMS-02 measurements 
. We remark that in this analyses it is crucial the use of the current AMS-02 data, including the recently published fluxes of Mg, Si and Ne~\cite{AMS_Ne}, thus providing accurate measurements of the fluxes of the main primary CR species producing B, Be and Li~\cite{Aguilar:2015ooa, aguilar2015precision, Aguilar2016, Aguilar:2017hno, Aguilar:2018keu, Aguilar:2018njt}.

In order to perform the fit of the ratios, a Markov chain Monte Carlo (MCMC) procedure, relying on Bayesian inference, has been implemented to get the probability distribution functions for a set of diffusion parameters ($D_0$, $V_A$, $\eta$ and $\delta$) to describe the data (in this case the AMS-02 data) and their confidence intervals. The technical details are presented in the next section.

\subsection{MCMC and minimization algorithm}
\label{sec:MCMC}

Bayesian inference is used to get the posterior probability distribution function (PDFs), $\mathscr{P}$, for a set of diffusion parameters to explain the AMS-02 data. To evaluate it, the prior PDF, $\Pi$, and the likelihood, $\mathscr{L}$, must be defined. These three terms are related by:
\begin{equation}
\mathscr{P}(\vec{\theta}|\vec{D}) \propto  \mathscr{L}(\vec{D}|\vec{\theta}) \Pi(\vec{\theta})
\label{eq:bayesian} 
\end{equation}
where $\vec{\theta} = \{\theta_1, \theta_2, ... , \theta_m\}$ is the set of parameters, $\vec{D}$ is the data set used (CR flux ratios measured by AMS-02) and $\mathscr{P}(\vec{\theta}|\vec{D})$ is the posterior PDF for the parameters. 

The posterior probability is calculated from equation~\ref{eq:bayesian} by a MCMC algorithm which uses, instead of the classical \textit{Metropolis-Hastings} algorithm, a modified version of the \textit{Goodman \& Weare} algorithm \cite{Goodman_Weare}. This method is directly implemented in the \textit{emcee} module of Python (see ref.  \cite{emcee} for full technical information).

\begin{table}[!t]
\centering
\resizebox*{0.37\columnwidth}{0.1\textheight}{
\begin{tabular}{|lc|}
  \multicolumn{2}{c}{\hspace{0.3cm}\large\textbf{ \ \ \ \ \ }} \\ \hline & \textbf{Parameters range}\\ 
  \hline
{$D_0$} ($10^{28} cm^{2}/s$) & [5.5 , 9.5] \\
{$v_A$} ($km/s$) & [0 , 40] \\
{$\eta$}    & [-3. , 0.] \\       
{$\delta$}    & [0.33, 0.53] \\     
\hline
\end{tabular}}
\caption{Parameter space considered in the analyses.}
\label{tab:ranges}
\end{table}

In this procedure 
a ``grid'' of simulations is built for $\sim 10000$ regularly spaced combinations of parameters in order to interpolate them. The range of parameters used to build this grid is shown in table~\ref{tab:ranges}. For each of these combinations, we compute the flux ratios among different CR species and carry out interpolations for other combinations of propagation parameters by using the python tool \textit{RegularGridInterpolator} from the \textit{Scipy} module. In other words, we create a sort of four-dimensional ``matrix'' (a dimension for each diffusion parameter) and, to each of the points of the matrix, we associate a vector containing the simulated flux ratios at different energies. The diffusion equation is integrated in 171 points with equal spacing in a logarithmic scale, from $10 \units{MeV/n}$ to $\sim100 \units{TeV/n}$.
The errors coming from this interpolation were estimated comparing different simulated fluxes of different nuclei with the interpolated fluxes, obtaining errors always much smaller than 1\%. Errors in the interpolation of fluxes in different energies are completely negligible. 


The likelihood $\mathscr{L}$ is set to be a Gaussian function (as in many other studies, e.g.  \cite{Trotta:2010mx} or \cite{Niu:2018waj}):

\begin{equation}
\mathscr{L}(\vec{D}|\vec{\theta}) = \prod_{i,j}\frac{1}{\sqrt{2\pi\sigma_{i,j}^2}} \exp \left[- \frac{(\Phi_{i,j}(\vec{\theta}) - \Phi_{i,j, data})^2}{2\sigma^2_{i,j}} \right]
\label{eq:likelihood} 
\end{equation}
where $\Phi_{i,j}(\vec{\theta})$ is the flux ratio computed in the simulation, $\Phi_{i,j, data}$ is the corresponding flux ratio measured by AMS-02 and $\sigma_{i,j}$ is its associated error for the $i$-th flux ratio and for the $j$-th energy bin. 
In this approach we do not take into account possible correlations among data, as the public data from the AMS-02 Collaboration do not include the full covariance matrix~\cite{Weinrich:2020cmw, Heisig_correl}. Therefore, the experimental uncertainties are taken to be the sum in quadrature of statistical and systematic errors reported by the AMS-02 Collaboration.
The prior PDF is defined as a uniform distribution for all the parameters:

\begin{equation}
\Pi(\vec{\theta}) = \left\{
\begin{array}{ll}
\prod_{i} \cfrac{1}{\theta_{i, max} - \theta_{i, min}}  &  
\text{for $\theta_{i, min} < \theta_i < \theta_{i, max}$} \\
 0    & \text{elsewhere} \\ 
\end{array}
\right.
\label{eq:prior} 
\end{equation}

Then, in order to take into account cross sections uncertainties, we make use of a nuisance parameter for each of the secondary CRs analysed (i.e. B, Be and Li), which enables a renormalization (or scaling) of the parametrization used for their production cross sections. We argue that applying evenly this scale factor to every isotopic species is a simplification, but given that there is no isotopic data on CR spectra above $10 \units{GeV/n}$ this scaling can not be precisely applied for individual isotopes. The application of this scaling on the cross sections parametrizations can lead to a subtle change in the energy dependence of the low energy part ($\sim$ below $1\units{GeV/n}$) of the spectra of B, Be and Li. We highlight that, even though our determination of the propagation parameters focuses on AMS-02 data, our predictions on isotopic fluxes seem to be also compatible with the isotopic data from the ACE/CRIS experiment~\cite{Fu_2019} (although at the level of $2\sigma$ for Li isotopes) and from a balloon borne experiment ~\cite{1979ICRC....1..389W}, at energies around $100\units{MeV/n}$.
The associated nuisance term is usually defined as (eq. 4 of ref.~\cite{Weinrich:2020cmw}):
\begin{equation}
\mathscr{N}_X = \sum_i\frac{\left(y_{i,X} - \hat{y}_{i,X}\right)^2}{\sigma_{i,X}^2}
\label{eq:nuisance} 
\end{equation}
where $\hat{y}_{i,X}$ is the experimental cross section value at the $i$-th point of energy and $\sigma_{i,X}$ its associated experimental uncertainty referred to the secondary CR species $X$, that can be B, Be or Li. The term $y_{i,X}$ is defined as $y_{i,X} \equiv \mathcal{S}_X \cdot \hat{y}_{i,X}$, where $\mathcal{S}_X$ is the nuisance parameter to be adjusted, referred to the species X, and independent of energy. 

As the nuisance parameters are independent of energy, eq. \ref{eq:nuisance} can be rewritten as:  
\begin{equation}
\mathscr{N}_X = (\mathcal{S}_X - 1)^2 
\sum_i \left( \frac{\hat{y}_{i,X}}{\sigma_{i,X}}\right)^{2} = 
(\mathcal{S}_X - 1)^2 N \left\langle (\hat{y}/\sigma)^{2} \right\rangle_X
\label{eq:nuisance2}
\end{equation}
Here $N$ stands for the number of experimental data points and $\langle \sigma/\hat{y}\rangle_X$ is the average relative error for a given channel of production of $X$. Given the lack of cross sections experimental data, we use the reaction channels with $^{12}$C and $^{16}$O as projectiles. Then, the final expression for the nuisance term associated to a secondary CR species $X$ is:
\begin{equation}
    \mathscr{N}_X = (\mathcal{S}_X - 1)^2 \sum_{X'}(N_{C \rightarrow X'} \langle (\hat{y}/\sigma)^{-2} \rangle_{C \rightarrow X'} + N_{O \rightarrow X'} \langle (\hat{y}/\sigma)^{-2} \rangle_{O \rightarrow X'})
    \label{eq:myNuisance}
\end{equation}
where $X'$ stands for all possible isotopes of the CR species $X$ (for example, for B it means the sum for the isotopes $^{10}$B and $^{11}$B). The nuisance terms are therefore originated from Gaussian distributions with mean 1 and sigma given by $\langle \sigma/\hat{y} \rangle$ (these relative uncertainties are $>20\%$, as shown in ref.~\cite{Luque:2021joz}), and can be interpreted as penalty terms that prevent the scaling factors to be far from 1.


With this strategy we are including the normalization uncertainty of the cross sections in the analysis, thus enabling the cross sections to be scaled up or down, with each secondary being more penalized as the scaling associated to its production cross section deviates from 1. The effect of scaling all the production cross sections causes the fluxes to be scaled by the same factor. Therefore the fluxes of the CR species $X$ in the likelihood will be scaled of a factor $\mathcal{S}_X$. These changes in the inclusive cross sections of B, Be and Li production result in small changes of the inelastic (often called fragmentation) cross sections. We have checked that the impact of these changes in the energy dependence of the flux ratios among secondary CRs is negligible and we estimate that the uncertainty associated to this effect should be of, at most, $1\%$. Inelastic cross sections are a subdominant term in the propagation equation and the overall variation of the inelastic cross sections obtained in this work is very small given that most of the inelastic reactions produce He and protons. 

Nevertheless, when studying each ratio independently, the cross sections uncertainties are mostly absorbed by the normalization of the diffusion coefficient, as also reported in ref.~\cite{Weinrich:2020cmw}, which means that $\mathcal{S}_X$ will remain equal to 1 and no conclusions on these uncertainties can be obtained.
Combining the secondary-to-primary ratios of different species allows accessing this information. In order to include the full correlation between cross sections and CR fluxes, the secondary-over-secondary ratios are included in this combined analysis. As we have seen, they are a perfect tool to study cross sections parametrizations with very small uncertainty at high energies. Therefore including these ratios in the analysis can make the determination of the diffusion coefficient more robust, providing information about the correlations between the secondary species. In this way, we use the likelihood function defined using the secondary-over-secondary ratios above $30 \units{GeV}$, since, at these energies, their spectra is mostly dependent on the ratio of the cross sections of production of the secondary CRs involved.

In conclusion, the logarithm of the likelihood function used to carry out the combined analysis of Li, B and Be is given by:
\begin{equation}
     \ln \mathscr{L}^{Total} =  \sum^{Li,Be,B/(C,O,Li,Be,B)}_F \ln(\mathscr{L}(F)) + \frac{1}{2}\sum^{B, Be, Li}_X \mathscr{N}_X
    \label{eq:mylikelihood}
\end{equation}
where $F$ indicates the flux ratios (six secondary-over-primary ratios and three secondary-over-secondary ratios) and $X$ indicates the cross sections channels included (those coming from $^{12}$C and $^{16}$O projectiles only). 


\section{Results}
\label{sec:results}

The algorithm explained above has been applied to find the optimal diffusion parameters that provide the best fit of the spectra of each of the ratios of B, Be and Li to C and O (B/C, B/O, Be/C, Be/O, Li/C, Li/O) independently, for both diffusion models (eqs.~\ref{eq:sourcehyp} and~\ref{eq:breakhyp}). In section~\ref{sec:Standard_results}, we report the results of this analysis for each of flux ratios independently. Additionally, in this section we report the results for a combined analysis of the spectra of the secondary CRs B, Be and Li. In this analysis we combine the studied secondary-to-primary flux ratios and the secondary-over-secondary ratios Be/B, Li/B and Li/Be (at $E > 30 \units{GeV}$, in order to avoid the uncertainties at low energy, largely studied in~\cite{Luque:2021joz}) as well as the nuisance parameters accounting for uncertainties related to the normalization of the cross sections parametrizations. 

Then, in section~\ref{sec:CXSanal}, we employ another analysis in order to consider, in a different way, the uncertainties in the normalization of the cross sections parametrizations and obtain the best-fit diffusion parameters from the secondary-to-primary flux ratios. This strategy consist of re-scaling the cross sections of production of B, Be and Li by reproducing the high energy part of the spectra of the secondary-to-secondary flux ratios Be/B, Li/B and Li/Be, provided that this part of their spectra is intimately related to their cross sections and rather independent of the diffusion parameters used, as discussed in~\cite{Luque:2021joz}. After re-scaling the spallation cross sections used, we apply the MCMC analysis to each of the secondary-to-primary ratios. In this way, we can compare individually the parameters inferred from each secondary CR after ``correcting'' their cross sections. Although similar analysis have been presented in the past based on refitting the cross sections parametrizations using updated cross sections measurements (see, e.g., ref.~\cite{Tomassetti:2015nha}), the novelty here consists of renormalizing the cross sections parametrizations from a combination of the high energy part of the secondary-to-secondary spectra, taking the minimum possible rescaling which is able to simultaneously reproduce these ratios. In fact, this strategy has been successfully implemented for the study of antiprotons in ref.~\cite{Luque:2020crb}, where the author shows that, considering the scale factors calculated in this way, the antiproton-to-proton spectrum from the AMS-02 experiment can be reproduced with a high precision.

These analyses are carried out using two different spallation cross sections parametrizations: the {\tt DRAGON2} cross sections\footnote{Publicly available as ASCII files at \url{https://raw.githubusercontent.com/cosmicrays/DRAGON2-Beta\_version/master/data/crxsecs\_fragmentation\_Evoli2019\_cumulative\_modified.dat} }~\cite{Luque:2021joz, DRAGON2-2, Evoli:2019wwu} and the {\tt GALPROP}~\footnote{Available as ASCII files at \url{https://dmaurin.gitlab.io/USINE/input\_xs\_data.html\#nuclei-xs-nuclei}}~\cite{GALPROPXS, GALPROPXS1} cross sections parametrizations. 
It must be stressed here that the halo size values were determined in an independent way, from the $^{10}Be/^{9}Be$ flux ratios (which are roughly independent of the propagation parameters), as it is shown in section 5 of~\cite{Luque:2021joz}. Interestingly, the halo size values obtained from both parametrizations are very similar and compatible within $1\sigma$ statistical uncertainty: $6.76 \units{kpc}$ and $6.93 \units{kpc}$ for the {\tt DRAGON2} and {\tt GALPROP} parametrizations, respectively.

\subsection{Standard analysis}
\label{sec:Standard_results}
\begin{figure}[!t]
\centering
\includegraphics[width=\textwidth, height=0.55\textheight]{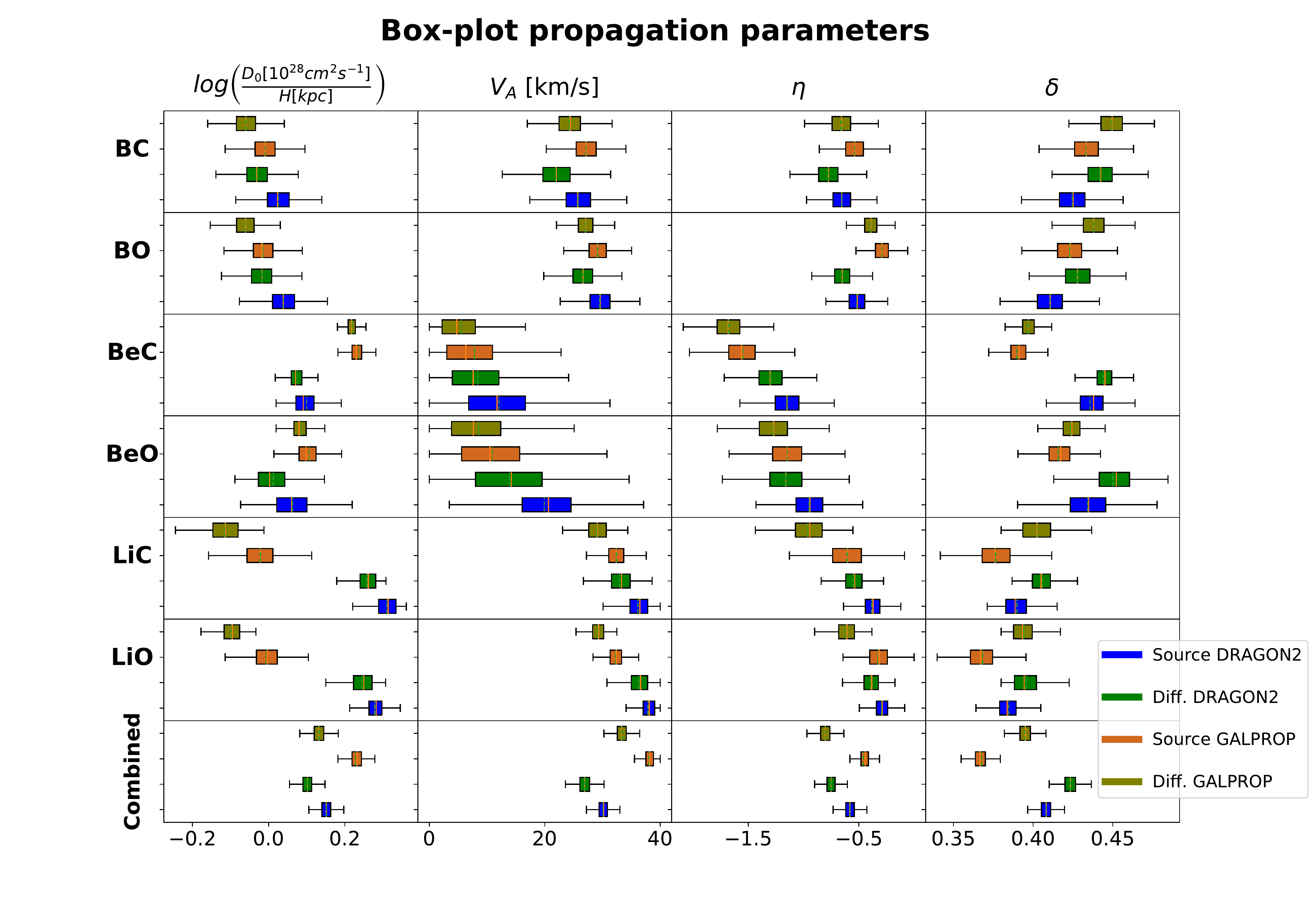}
\caption{Box-plots representing the PDFs resulting from the MCMC algorithm used to find the best propagation parameters to describe AMS-02 results with diffusion coefficient parametrizations of eq.~\ref{eq:sourcehyp}, and with the {\tt DRAGON2} and {\tt GALPROP} cross sections. The orange solid lines within each box indicate the medians of the corresponding PDFs, while the dashed cyan lines indicate the mean values; the colored boxes correspond to the interquartile ranges (IQR) from the $25$th (Q1) to the $75$th (Q3) percentile; finally, the black lines correspond to the range $[Q1-1.5IQR, Q3+1.5IQR]$ which, in case of gaussian PDFs contains the $99.3\%$ of the probability.}
\label{fig:boxplot_Standard}
\end{figure}

The results of the standard analyses are summarized in Figure~\ref{fig:boxplot_Standard}.
The posterior PDF of each parameter is usually very close to a Gaussian distribution function, although some of the PDFs obtained from the combined analyses of the {\tt GALPROP} cross sections with the source hypothesis of the diffusion coefficient exhibit asymmetric tails.

An explicit summary of the median values and of the ranges of propagation parameters within the 68 and 95\% of their PDFs is given in form of tables (tables~\ref{tab:PDF_Combined},~\ref{tab:PDF_single_Source} and~\ref{tab:PDF_single_Diff}) in the appendix~\ref{sec:appendixA} for each of the cross sections and diffusion coefficient parametrizations. In addition, the corner plots with the PDFs for the combined analyses are also shown in the appendix, for both the source and diffusion hypotheses.

Very similar results (always within 2$\sigma$) are found for the propagation parameters analyzed in both hypotheses for the CR hardening. In general, the diffusion hypothesis favors smaller values of $D_0$, $V_A$ and $\eta$ and larger values of $\delta$. This is something expected, since the slope (controlled by the $\delta$ parameter) in the source hypothesis tends to be smaller, to compensate the hardening observed in CR spectra at the highest energies. Therefore, if $\delta$ is larger, $D_0$ must be smaller to balance this change (see the contour plots in appendix~\ref{sec:appendixA} where the correlations between parameters are explicitly shown). In the following, some important remarks are discussed for these results.

\subsubsection{Comparison among the trend of different ratios}

In general, the relative uncertainties in the determination of propagation parameters are larger in the $\eta$ and $V_A$ parameters,  for the independent analyses, mainly due to their degeneracy and the need of more data points at the low energy region.

The propagation parameters obtained for each pair of ratios involving the same secondary (e.g. B/C and B/O) are compatible with each other (although in the case of Be, for the {\tt GALPROP} parametrizations, we observe discrepancies up to the level of $2\sigma$). Small discrepancies can arise, mainly due to the production of secondary carbon, which is important at low energies.
    
Each of the ratios (also in the combined analysis and with both cross sections sets used) favors a negative value of $\eta$, theoretically motivated by dissipation of MHD waves~\cite{Ptuskin_2006} or appearance of non-resonant interactions at low energies~\cite{Reichherzer:2019dmb}. The Be ratios show clearly smaller values of this parameter for every cross sections set used, but this seems to be due to the halo size used in the simulations. The degeneracy between these two parameters makes one compensate the defects of the other, thus requiring much lower $\eta$ values. 

Interestingly, the value of $\delta$ is usually around 0.42-0.45 for the B and Be ratios (although the prediction of the Be/C ratio from the {\tt GALPROP} parametrization is lower by at least $0.02$ units) while a lower value is favored for the Li ratios (between 0.36 and 0.40). The fact that the Li spectrum is harder at high energies (i.e. smaller $\delta$ value) than those of other species was pointed out in ref.~\cite{Boschini2020} and lead the authors to suggest the presence of primary Li. Nevertheless, provided that the uncertainties related to the spallation cross sections for Li production are specially large, this is not clear at all (and a proof is the large difference in the inferred $D_0$ value between both cross sections parametrizations for the Li flux ratios).
    
The values of $V_A$ obtained are smaller than $30 \units{km/s}$ for the B ratios ($\sim 26\units{km/s}$), while the Li ratios yields larger values (around $35 \units{km/s}$). On the other hand, the $V_A$ values for Be ratios are much smaller, usually compatible with $V_A=0$. This is due to the degeneracy with the halo size employed in the simulations. Therefore, excluding the Be ratios, an average value of $V_A \sim 30 \units{km/s}$ is found, which seems to be in agreement with the expected values (see refs.~\cite{Spanier_2005, Spangler:2010nu, Lerche_2001}). 
    
\begin{figure}[t]
	\centering
	\includegraphics[width=0.49\textwidth, height=0.25\textheight]{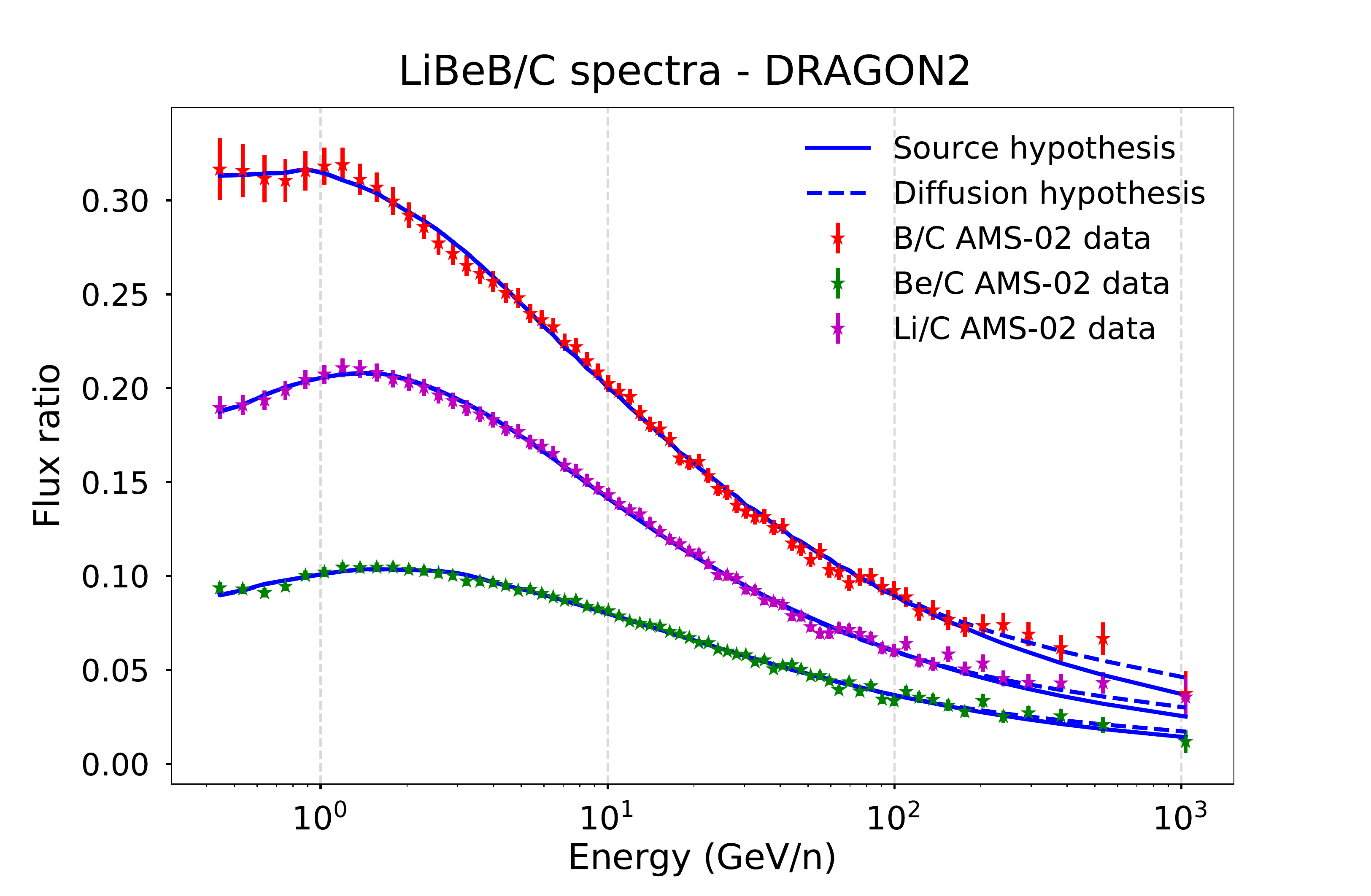}
	\hspace{-0.65cm}
	\includegraphics[width=0.49\textwidth, height=0.25\textheight]{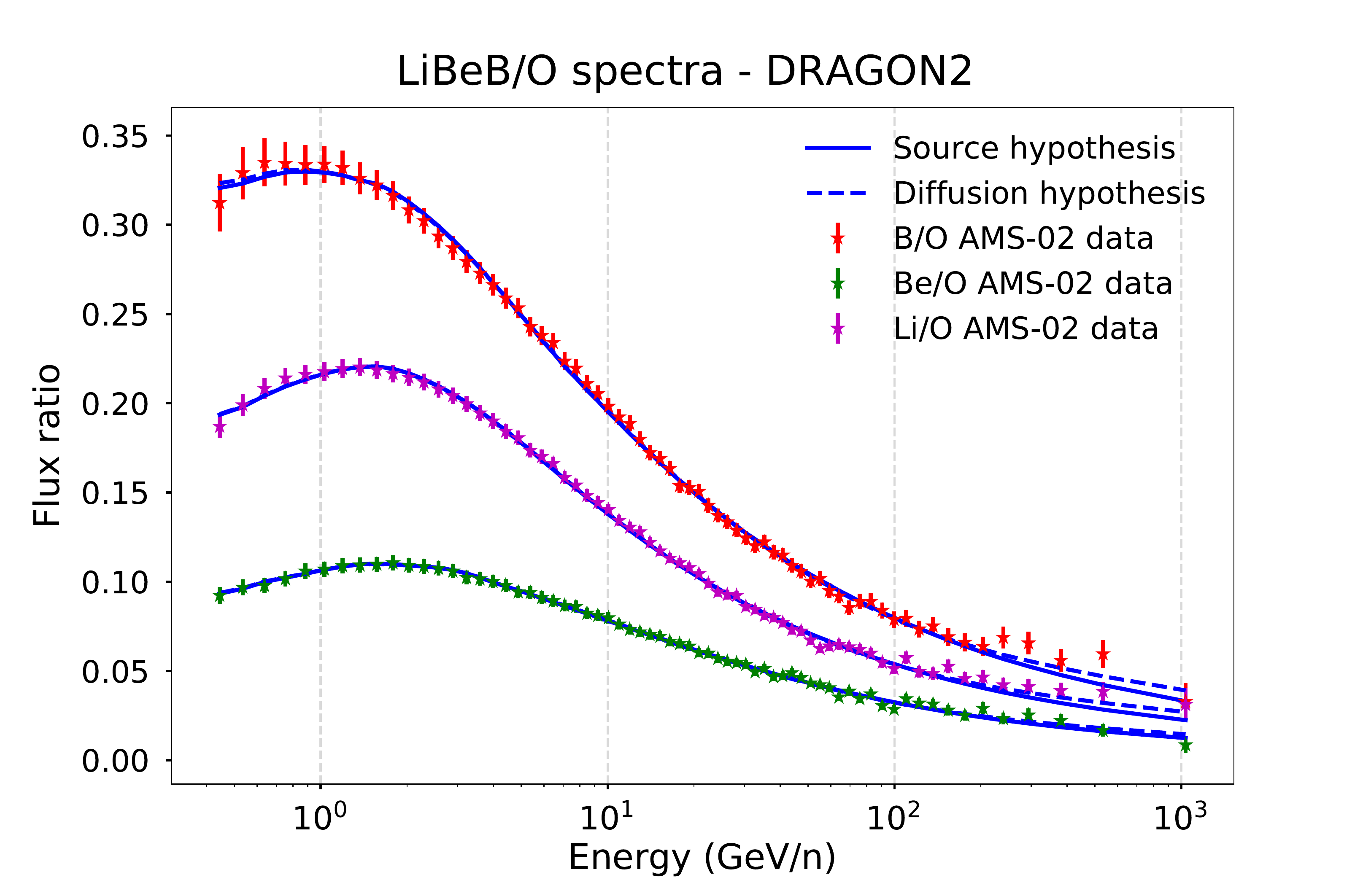}
	\caption{Predicted secondary-over-primary ratios with the propagation parameters determined independently in the MCMC analysis for the {\tt DRAGON2} cross sections parametrizations and for both hypothesis of the CR fluxes hardening. The parametrizations of the diffusion coefficients allow perfect reproduction of the shapes of the ratios. Fits with same quality are achieved for the {\tt GALPROP} cross sections too.}
	\label{fig:Single_ratios}
\end{figure}

Finally, as we see from Figure~\ref{fig:Single_ratios}, the secondary-over-primary flux ratios are perfectly reproduced when making each fit individually. This means that the degrees of freedom available in the parametrizations of the diffusion coefficient are enough to suitably match observations, although there is a clear improvement on the predictions when using the break in the diffusion (eq~\ref{eq:breakhyp}). On the other hand, it should be mentioned that the level of uncertainty appreciably changes when taking into account correlated errors from the AMS-02 experiment.


\subsubsection{Comparison among different cross sections sets}
Comparing both cross sections parametrizations by implementing a change in their normalization is, in fact, a way to see if the energy dependence of these parametrizations is compatible with the flux ratios of secondary CRs simultaneously. In general, the different cross sections give very similar predictions. One of the important differences are the $\eta$ and $V_A$ values obtained for the Be ratios with the {\tt DRAGON2} cross sections, which are larger 
than those obtained with {\tt GALPROP} cross sections. Remarkably, we also observe an important difference between both parametrizations in the predicted $\delta$ value obtained from the Be ratios. However, this might be also related to the degeneracy with the halo size value used.
    
The most meaningful difference is that, as discussed above, for both cross sections sets, the value of $\delta$ determined by the Li ratios is significantly smaller (often more than $2\sigma$) than the one determined by the other ratios. This points to the fact that, in both, parametrizations, the overall slope of the cross sections for Li production at high energies tends to be softer than it should be to be consistent with the $\delta$ values obtained from the B and Be analyses. Nonetheless, the best-fit $\delta$ value for the Li ratios with the {\tt DRAGON2} cross sections is closer to the value inferred from the B ratios, which might indicate that updating cross sections measurements can help to solve this issue.
    
\begin{figure}[!tb]
	\centering
	\textbf{\large{Cross sections scale factors}}
	\includegraphics[width=0.75\textwidth, height=0.4\textheight]{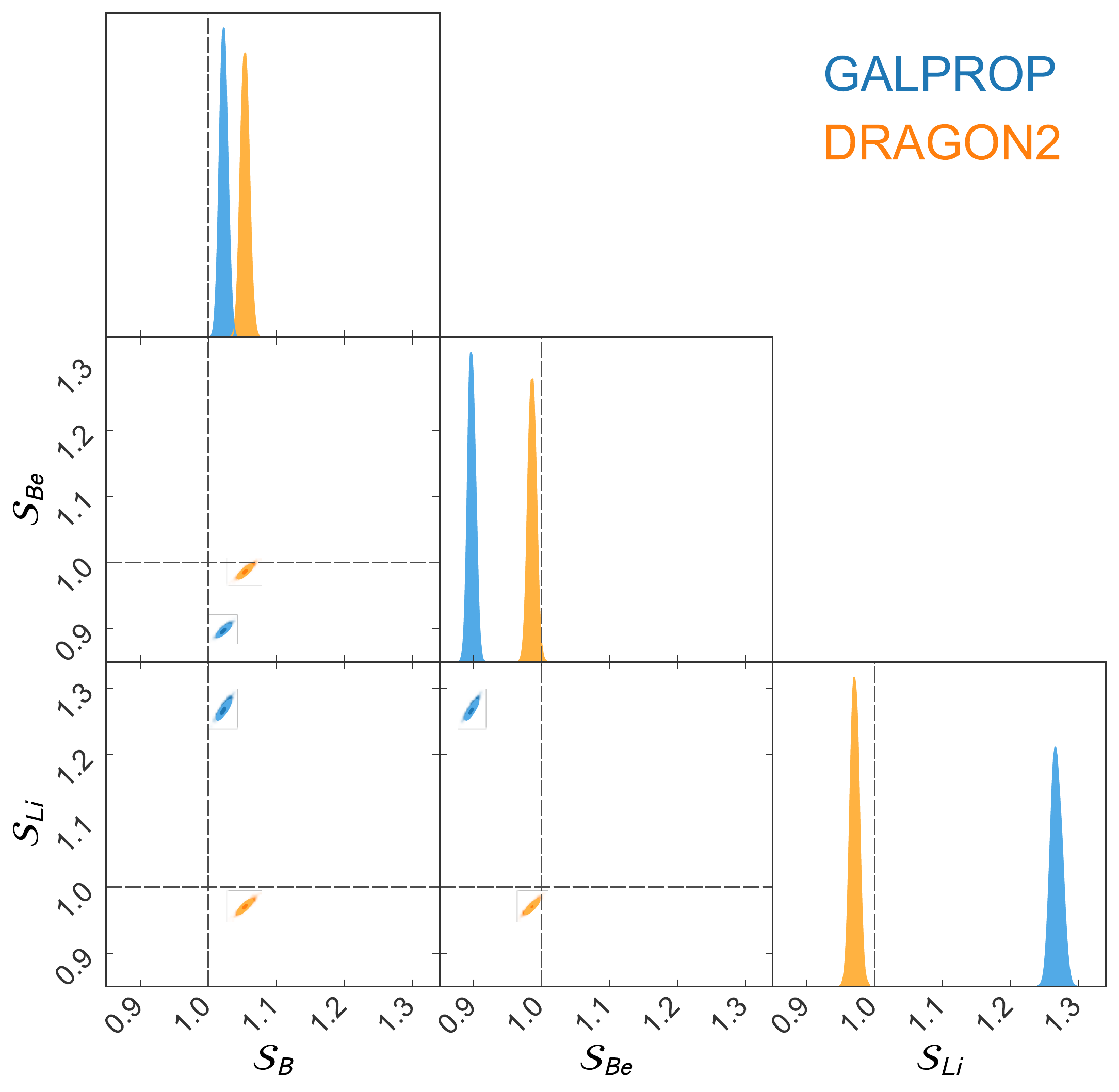}
	\caption{Corner plot of the scaling (nuisance) parameters determined by the MCMC combined analysis. A dashed line indicating scaling = 1 is added as reference for each of the panels.}
	\label{fig:Nuisance_combined}
\end{figure}

Finally, it is worth noticing that the $D_0$ values obtained are slightly different in the independent analyses of different secondary CRs, which mostly depends on the normalization of the spallation cross sections parametrizations in each channel. This is prevented by adding an overall scaling factor ($\mathcal{S}$) on the cross sections in the combined analysis. The relative differences of $D_0$ obtained and the secondary-over-secondary ratios for each of the cross sections sets reveal the relative degree of scaling on the production cross sections needed for each of the secondary CRs. These scaling factors, determined as nuisance parameters, are shown in Figure~\ref{fig:Nuisance_combined} and discussed in detail down below and in section~\ref{sec:CXSanal}.
    

\begin{figure}[!tb]
\centering
\hskip -0.15 cm	\includegraphics[width=0.53\textwidth, height=0.25\textheight]{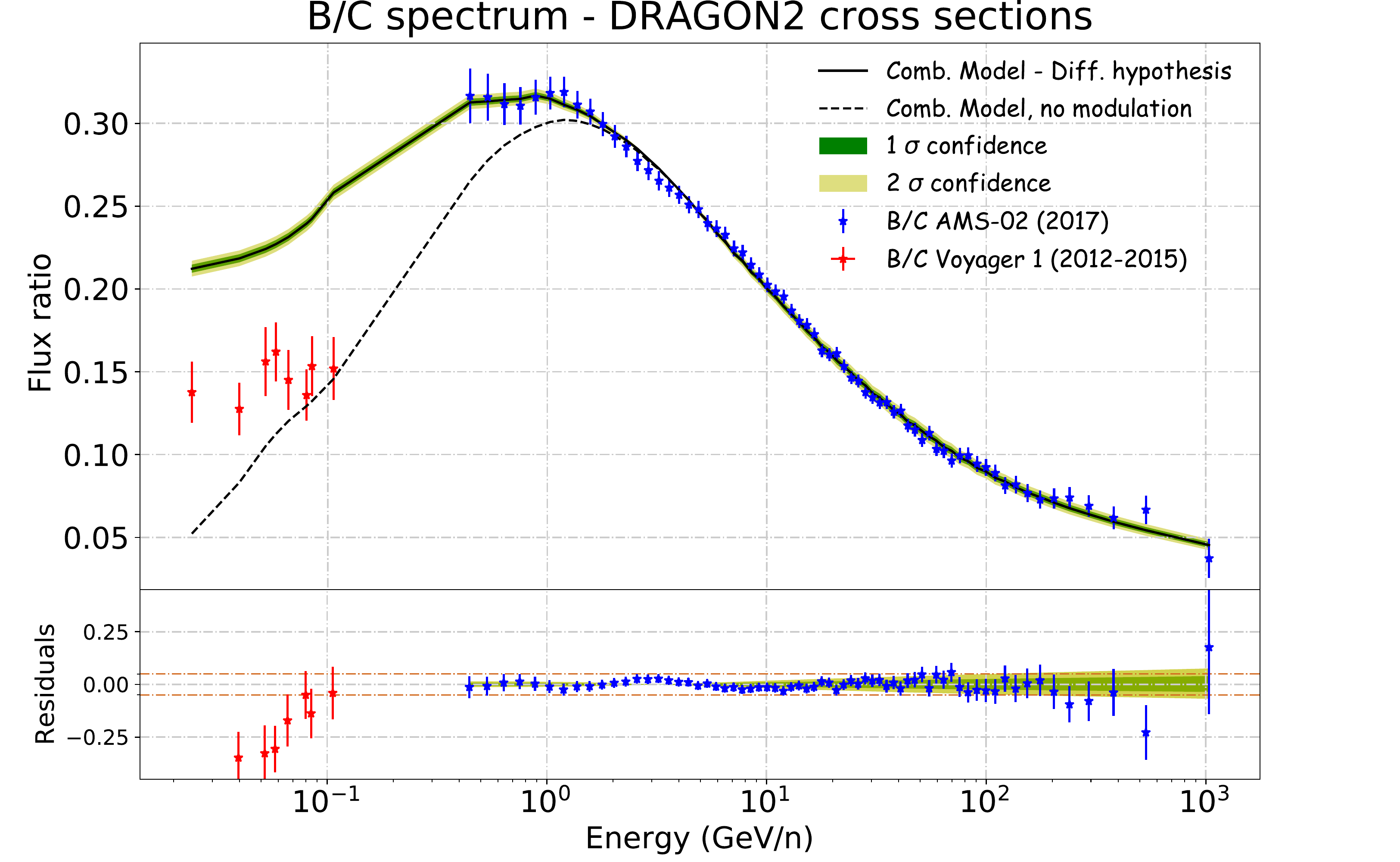}
\hspace{-0.5cm}
\includegraphics[width=0.53\textwidth, height=0.255\textheight]{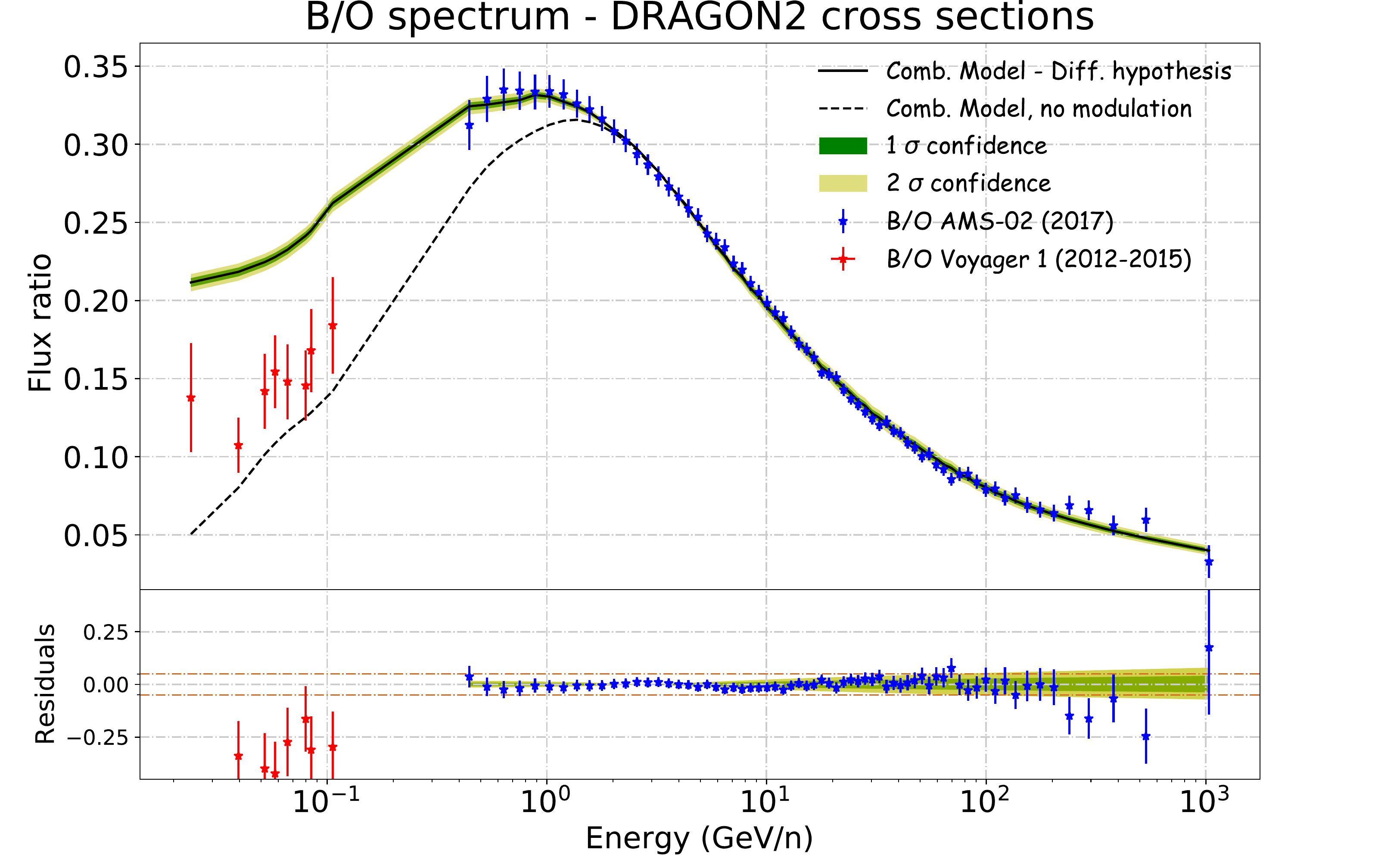}
\rightskip -0.6cm
	
\vspace{0.4cm}
	
\hskip -0.15 cm
\includegraphics[width=0.53\textwidth, height=0.255\textheight]{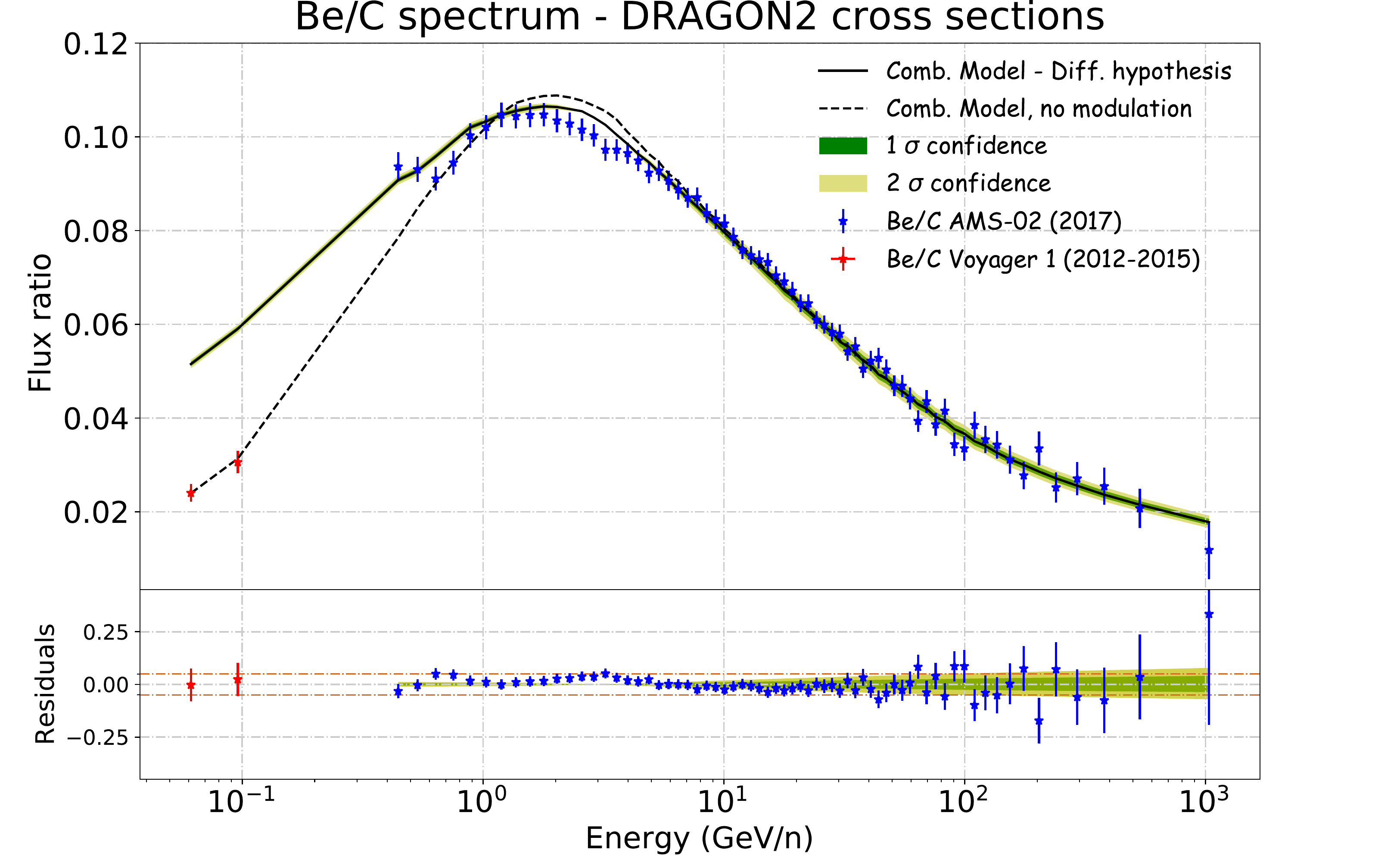}
\hspace{-0.5cm} \includegraphics[width=0.53\textwidth, height=0.255\textheight]{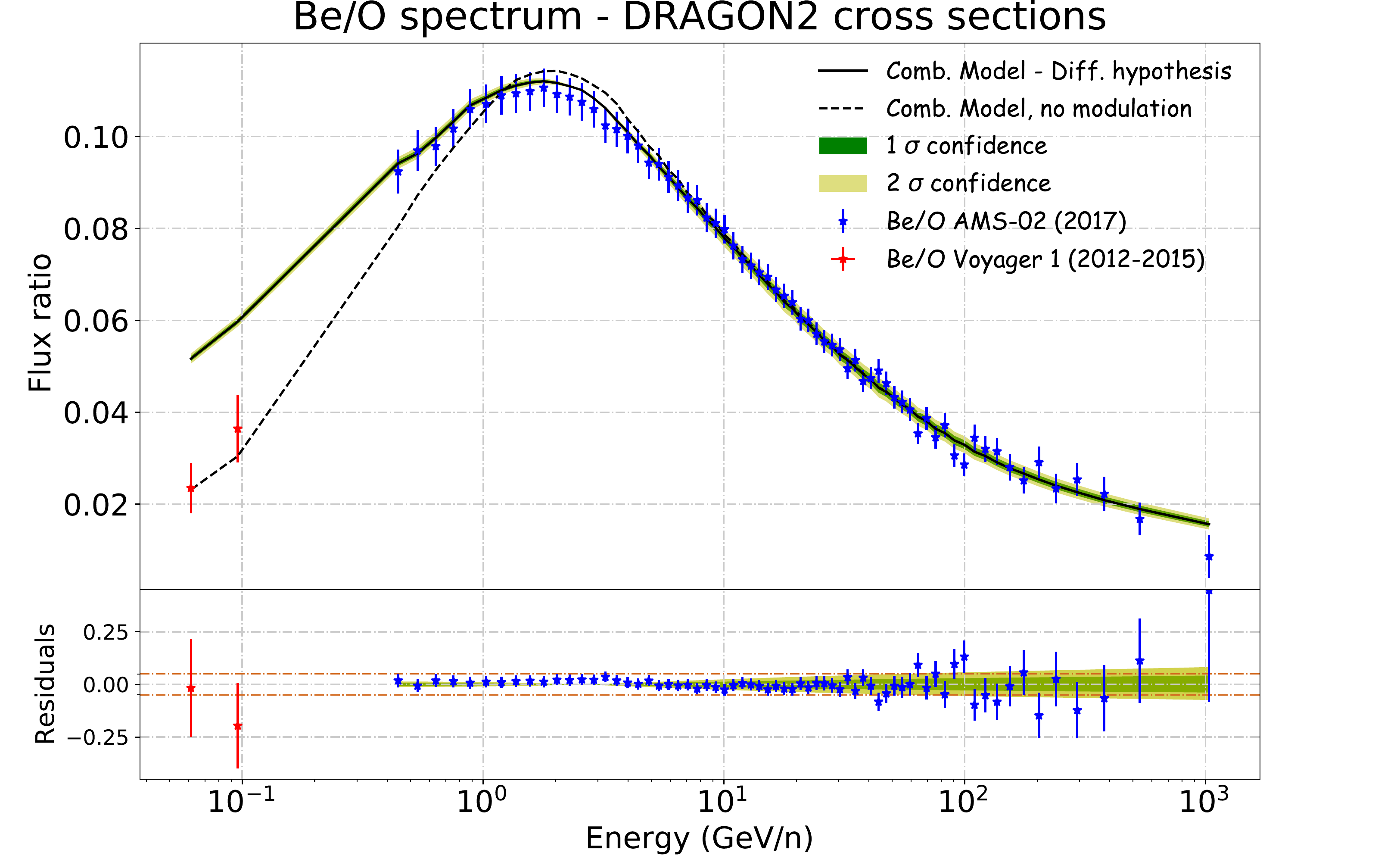}
\rightskip -0.6cm

\vspace{0.4cm}
	
\hskip -0.15 cm
\includegraphics[width=0.53\textwidth, height=0.255\textheight]{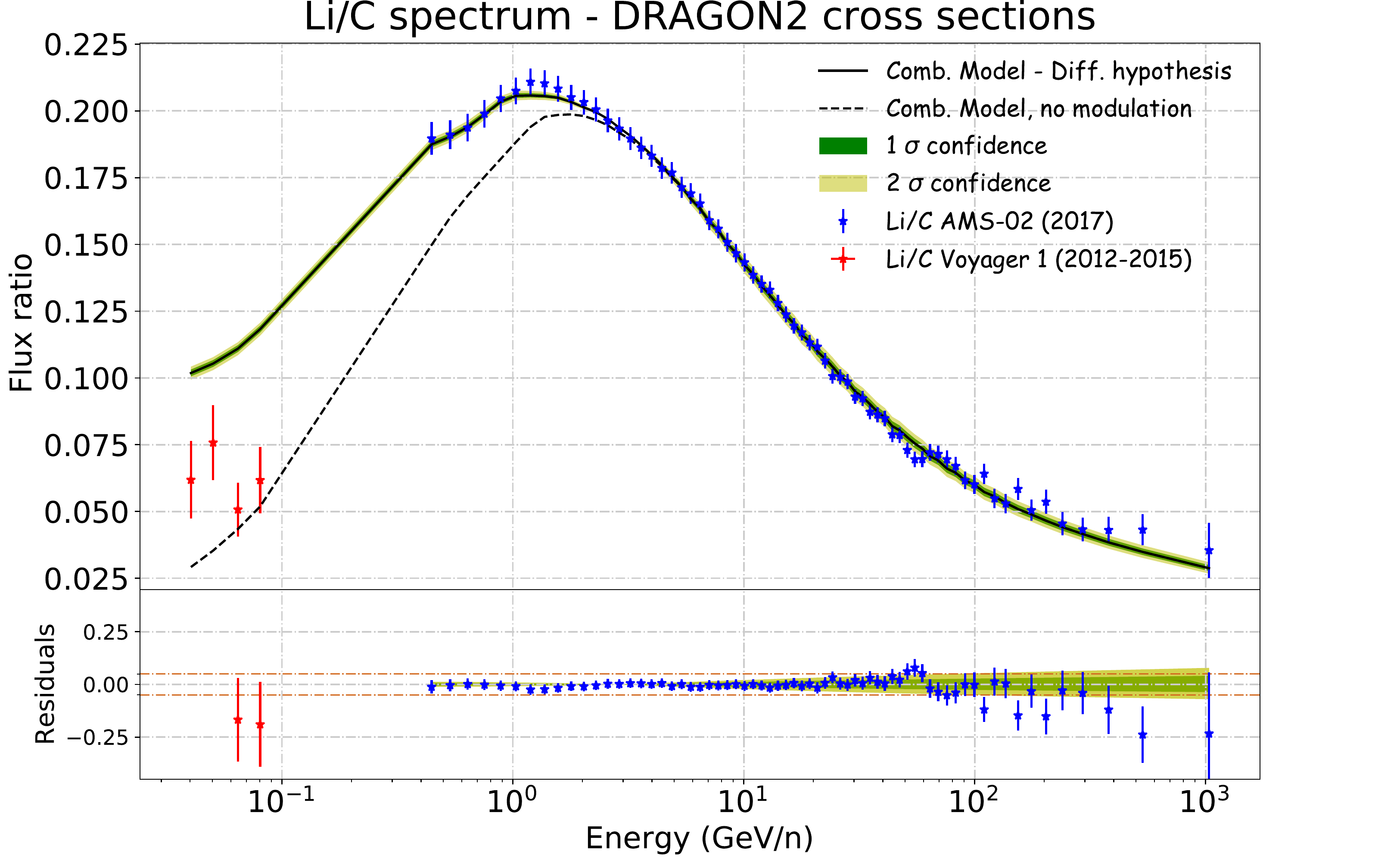}
\hspace{-0.5cm}
\includegraphics[width=0.53\textwidth, height=0.255\textheight]{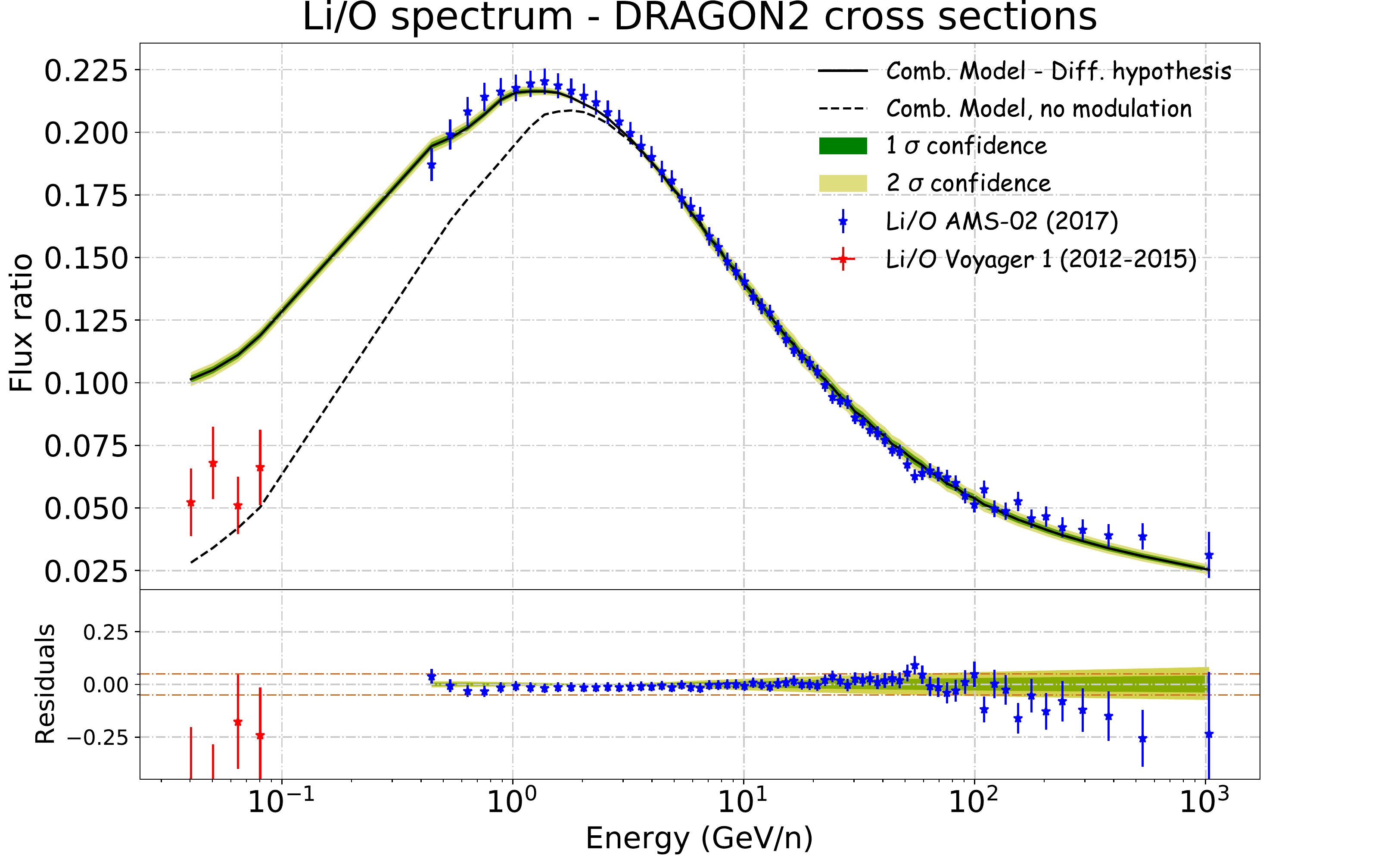}
\rightskip -0.6cm
\caption{Secondary-over-primary spectra (B/C, B/O, Be/C, Be/O, Li/C, Li/O) predicted with the parameters determined by the MCMC combined analysis for the {\tt DRAGON2} cross sections. The grid lines at the level of 5\% residuals are highlighted in the lower panel in a different color for clarity. In addition, the Voyager-1 data are also included for completeness.}
\label{fig:SecPrim_DRAGON2}
\end{figure}

\begin{figure}[!tb]
\centering
\hskip -0.15 cm \includegraphics[width=0.53\textwidth, height=0.255\textheight]{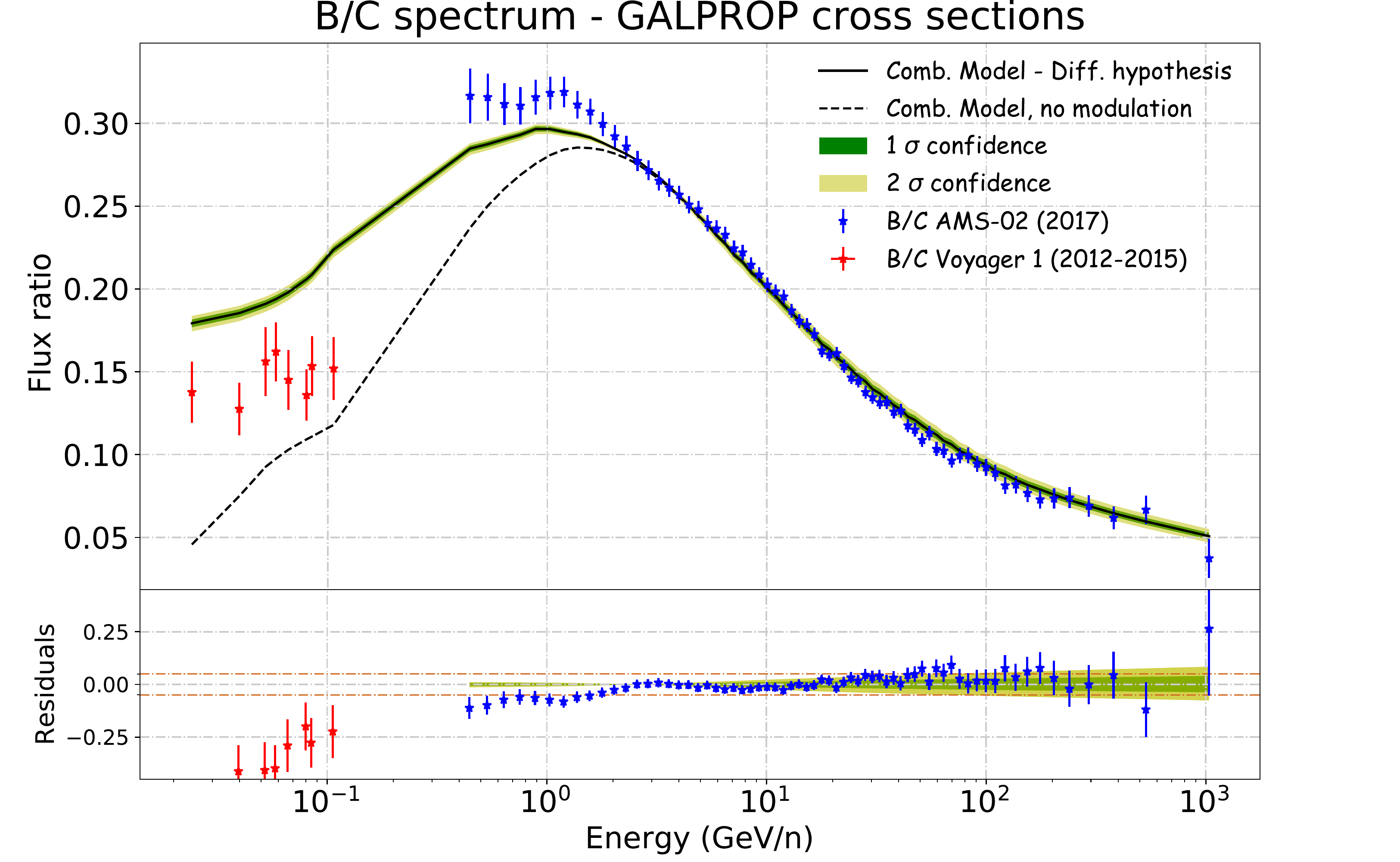}
\hspace{-0.5cm}
\includegraphics[width=0.53\textwidth, height=0.255\textheight]{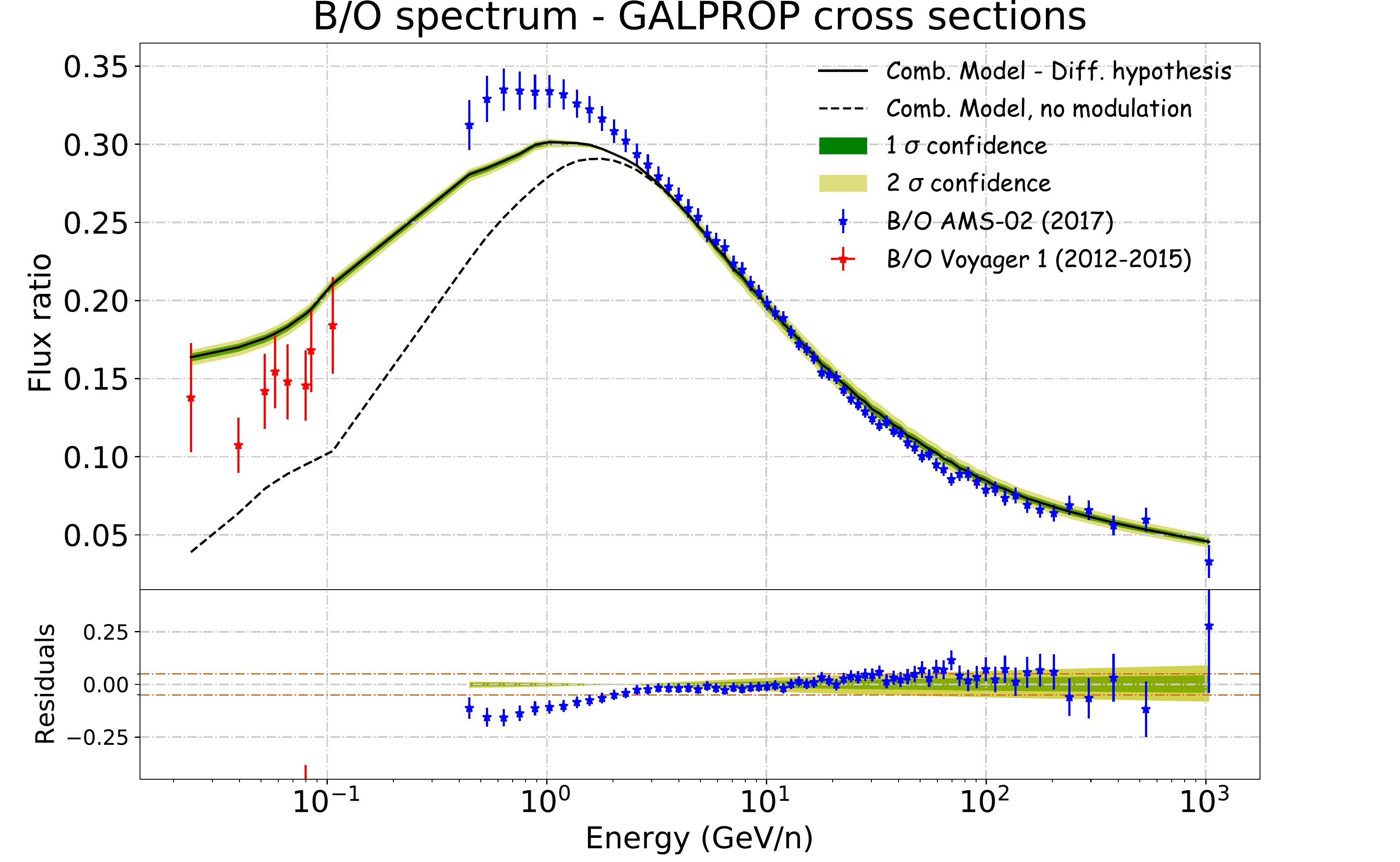}
\rightskip -0.6cm

\vspace{0.4cm}
	
\hskip -0.15 cm \includegraphics[width=0.53\textwidth, height=0.255\textheight]{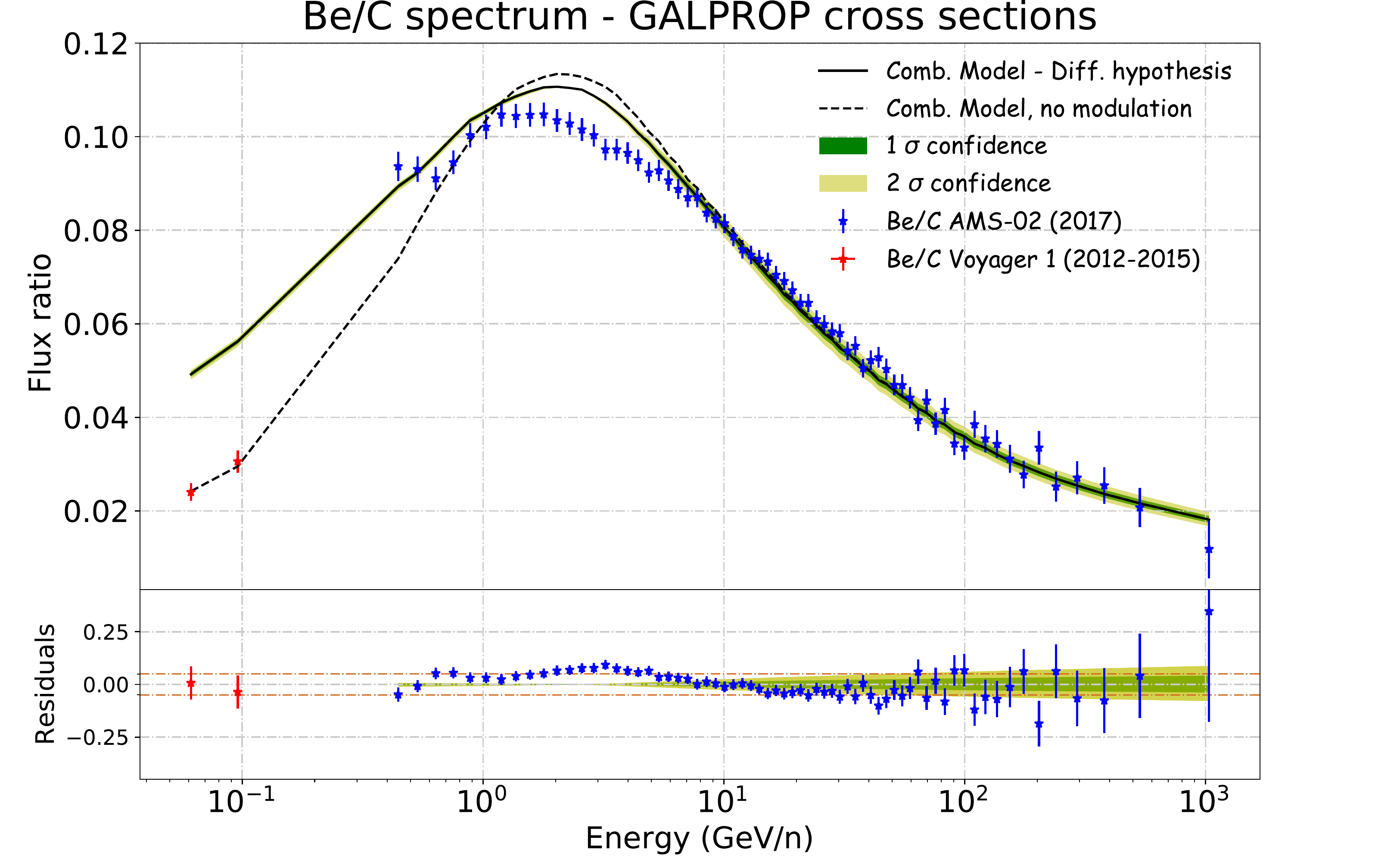}
\hspace{-0.5cm}
\includegraphics[width=0.53\textwidth, height=0.255\textheight]{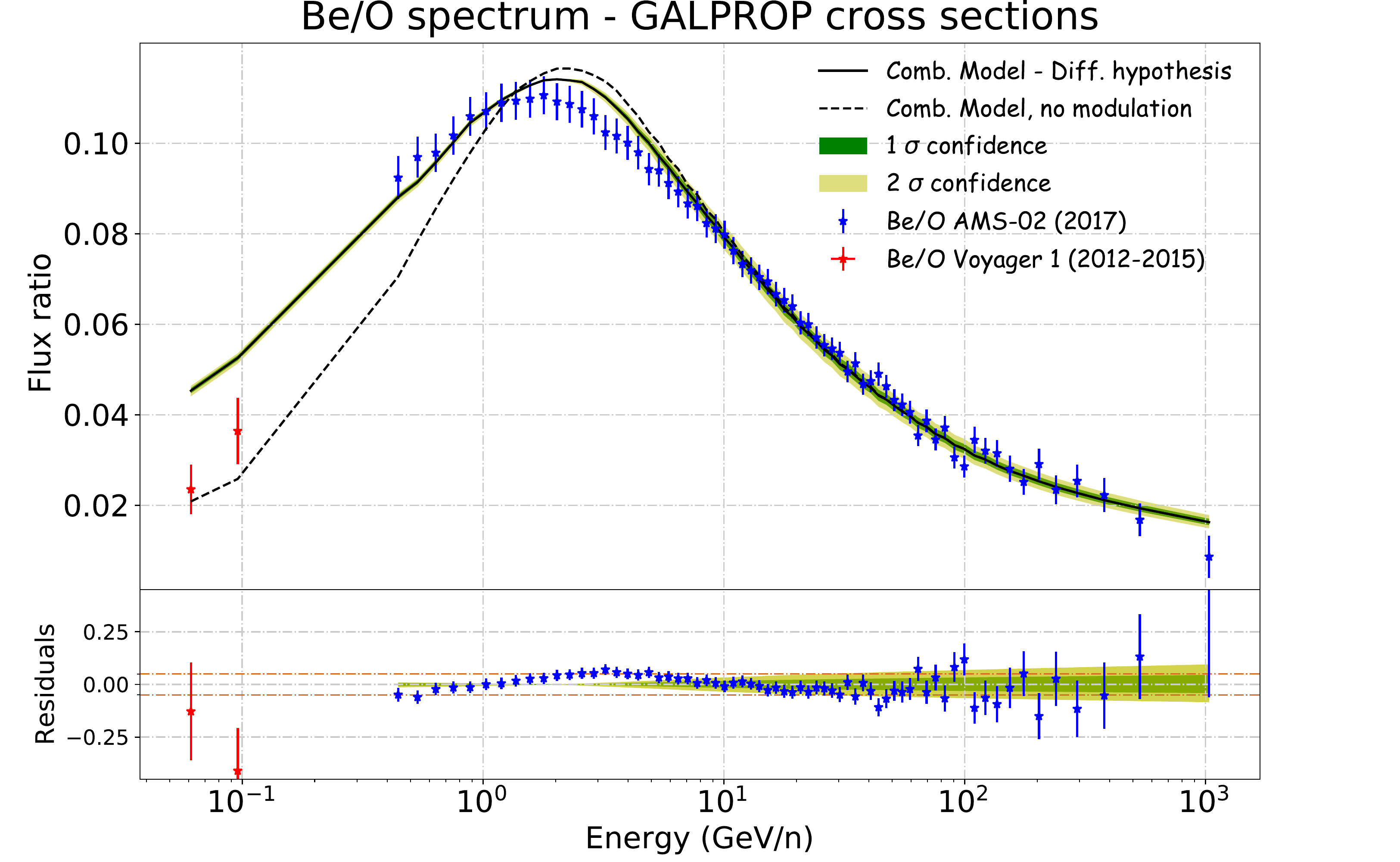}
\rightskip -0.6cm

\vspace{0.4cm}
	
\hskip -0.15 cm
\includegraphics[width=0.53\textwidth, height=0.255\textheight]{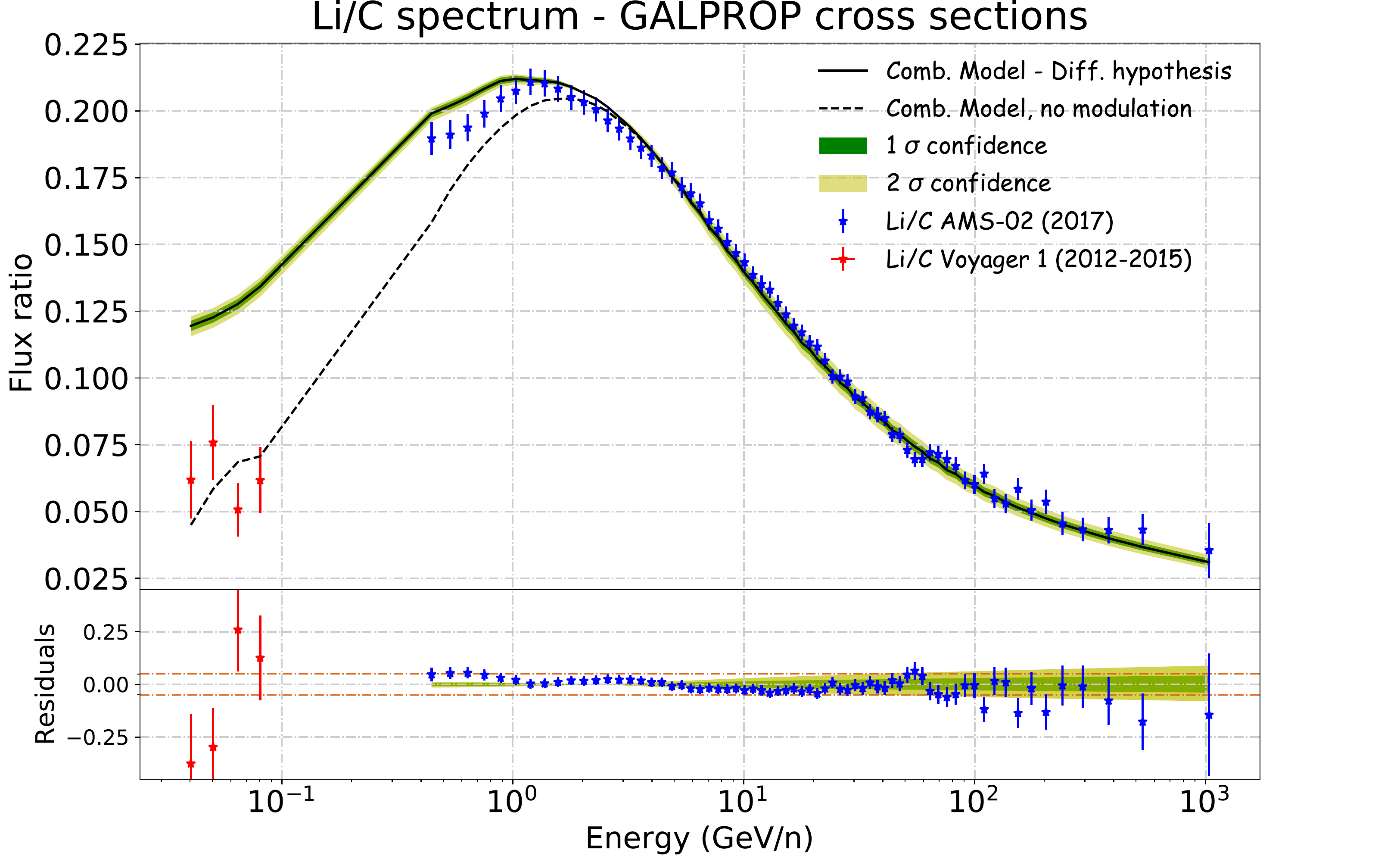}
\hspace{-0.5cm}
\includegraphics[width=0.53\textwidth, height=0.255\textheight]{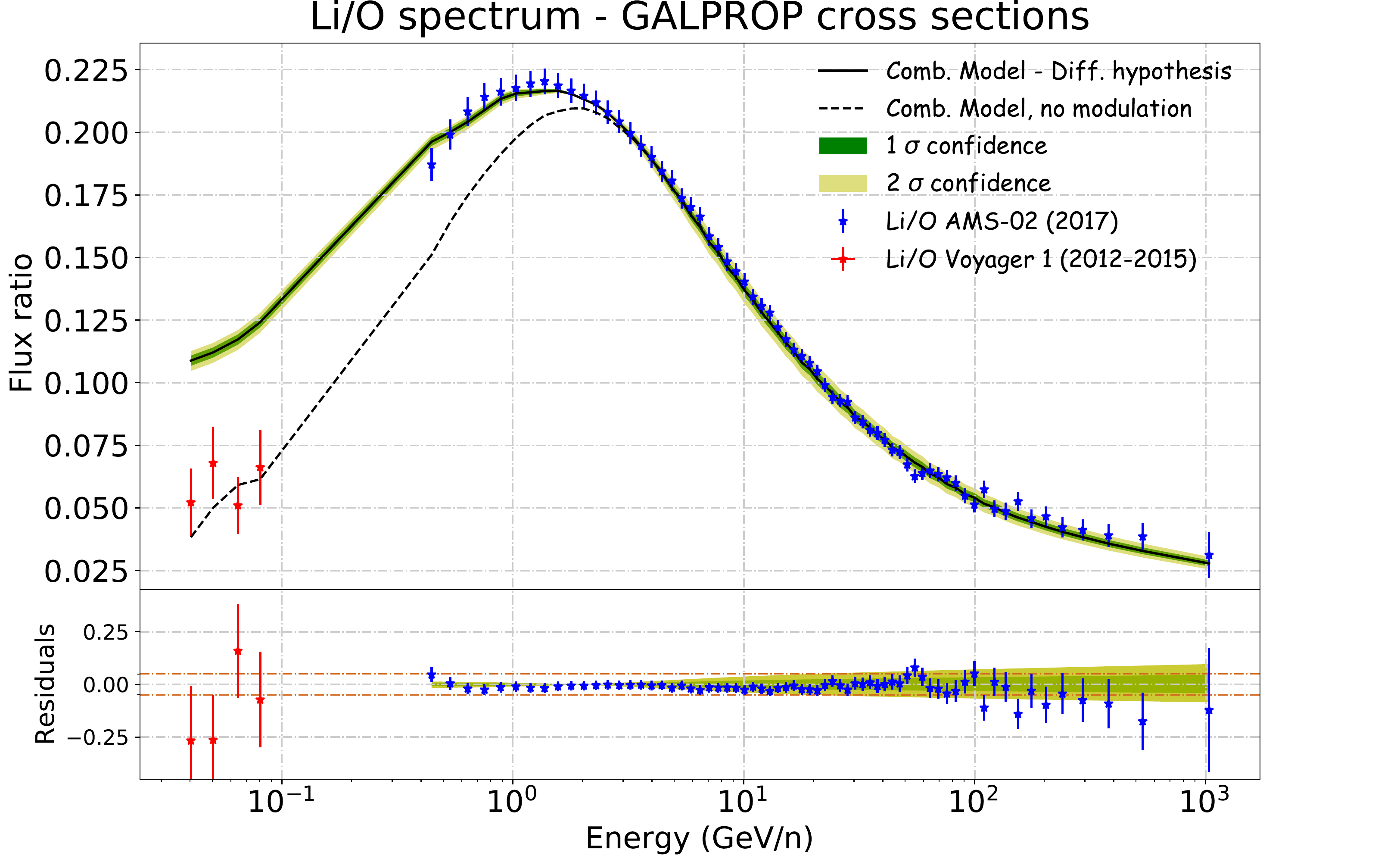}
\rightskip -0.6cm
\caption{Same as in Fig. \ref{fig:SecPrim_DRAGON2} but with the {\tt GALPROP} cross sections.}
	\label{fig:SecPrim_GALPROP}
\end{figure}

\subsubsection{Comparing combined vs independent analyses of the ratios}
The most remarkable fact seems to be that the values inferred from the combined analysis tend to 
be closer to those inferred from the Li ratios, leading to smaller $\delta$ values in both parametrizations and larger $V_A$ values for the {\tt GALPROP} cross sections.
This could imply that this analysis compensates possible deviations in the adjustment of the energy dependence of the cross sections of Li production with modifications of the diffusion parameters.
Only the {\tt DRAGON2} cross sections show a median value of $\delta$ higher than 0.40, but also in this case it is below the median for B and Be ratios. This is likely related to the lack of experimental data on cross sections of Li production (implying a poorer parametrization of the Li production cross sections) and may cause an important bias in our determination of the propagation parameters (and also in the determination of the nuisance parameters) in the combined analysis. In general, it seems that the parameters found in the combined and independent analyses are compatible for both parametrizations. While the best-fit parameters found in the independent analyses are similar for both cross sections parametrizations, the predicted best-fit value for $\delta$ significantly differ in the combined analyses of both parametrizations.

The results of the combined analysis are shown in Figs.~\ref{fig:SecPrim_DRAGON2} and~\ref{fig:SecPrim_GALPROP} for the {\tt DRAGON2} and {\tt GALPROP} cross sections, respectively. As we see, the results of the combined analysis with the {\tt DRAGON2} cross sections show an almost perfect fit of all the ratios, with the largest discrepancies for the Be ratios, which could be due to the halo size used. In turn, with the {\tt GALPROP} cross sections the predictions are noticeably discrepant for Be and B ratios at low energies, while only the Li ratios are well reproduced within experimental errors. This could be due to possible deviations in the adjustment of the shape describing the cross sections, as commented above, but other authors have solved this discrepancy by adding an extra source of lithium (related to Li production from novae)~\cite{Boschini2020}. 

From Figure~\ref{fig:Nuisance_combined} we see for the {\tt GALPROP} cross sections that, while the scaling factors for Li and Be are the largest ($\sim$26 and $10\%$ respectively), the scaling for B is the smallest ($1-2\%$ up) and it is even compatible with no scaling. However, the {\tt DRAGON2} cross sections parametrizations determine a scaling for B, Be and Li of $5\%$, $1\%$ and $3\%$ ($\pm$1\%), respectively.
   
To make a rough estimation of the total uncertainty in the determination of these scaling factors we must consider that they are mostly fixed by the high energy part of the secondary-over-secondary ratios. Thus, a $\sim3$\% uncertainty, due to the inelastic (fragmentation) cross sections, is reasonable for the flux of secondary CRs at high energies (see~\cite{Genoliniranking}). Moreover the uncertainties in the high energy part of these ratios, associated to the determination of the diffusion parameters, are $\sim (3 - 4)\%$, as shown in~\cite{Luque:2021joz}, due to the effect of multi-step spallation reactions. This makes the total uncertainties in the determination of the scaling factors to be at least of $\pm 5\%$ (assuming to be uncorrelated, i.e. summed in quadrature). On top of this, there is some additional uncertainty associated to the energy dependence of the cross sections parametrizations, which can affect the determination of the scaling factors by an additional $\sim \pm (3-5)$\%. Therefore, we estimate that the uncertainty in the determination of the scaling factors can be as high as $\sim \pm 7-8$\%, which is well below half of the average experimental uncertainty associated to the cross sections measurements~\cite{Luque:2021joz}. Furthermore, we remark that using a different gas distribution can affect the determination of these scaling factors, since up to a $5\%$ change has been observed in the secondary-to-secondary flux ratios above a few GeV/n.


\begin{figure}[!t]
\centering
\includegraphics[width=0.51\textwidth, height=0.26\textheight]{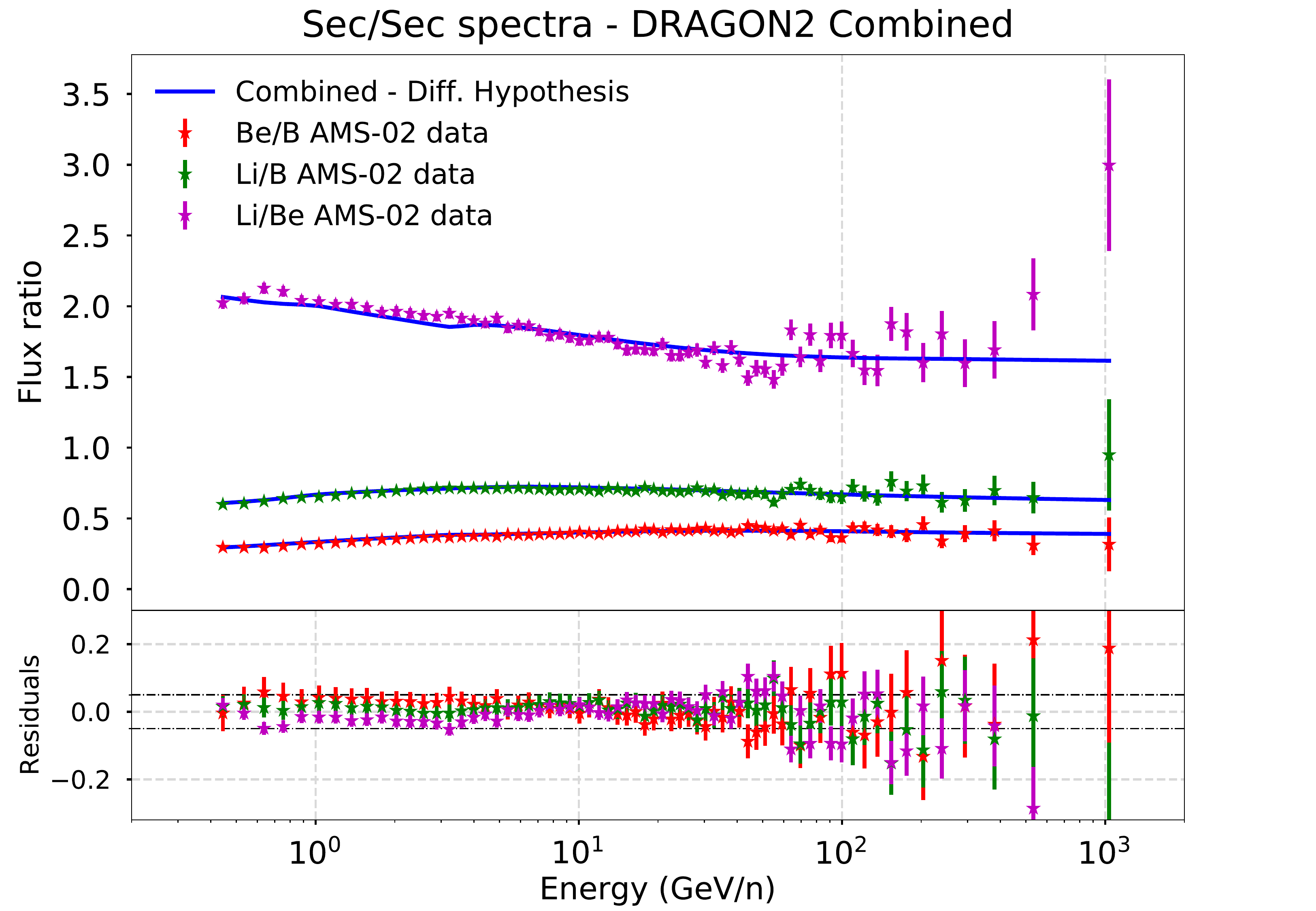} \hspace{-0.3cm}
\includegraphics[width=0.51\textwidth, height=0.26\textheight]{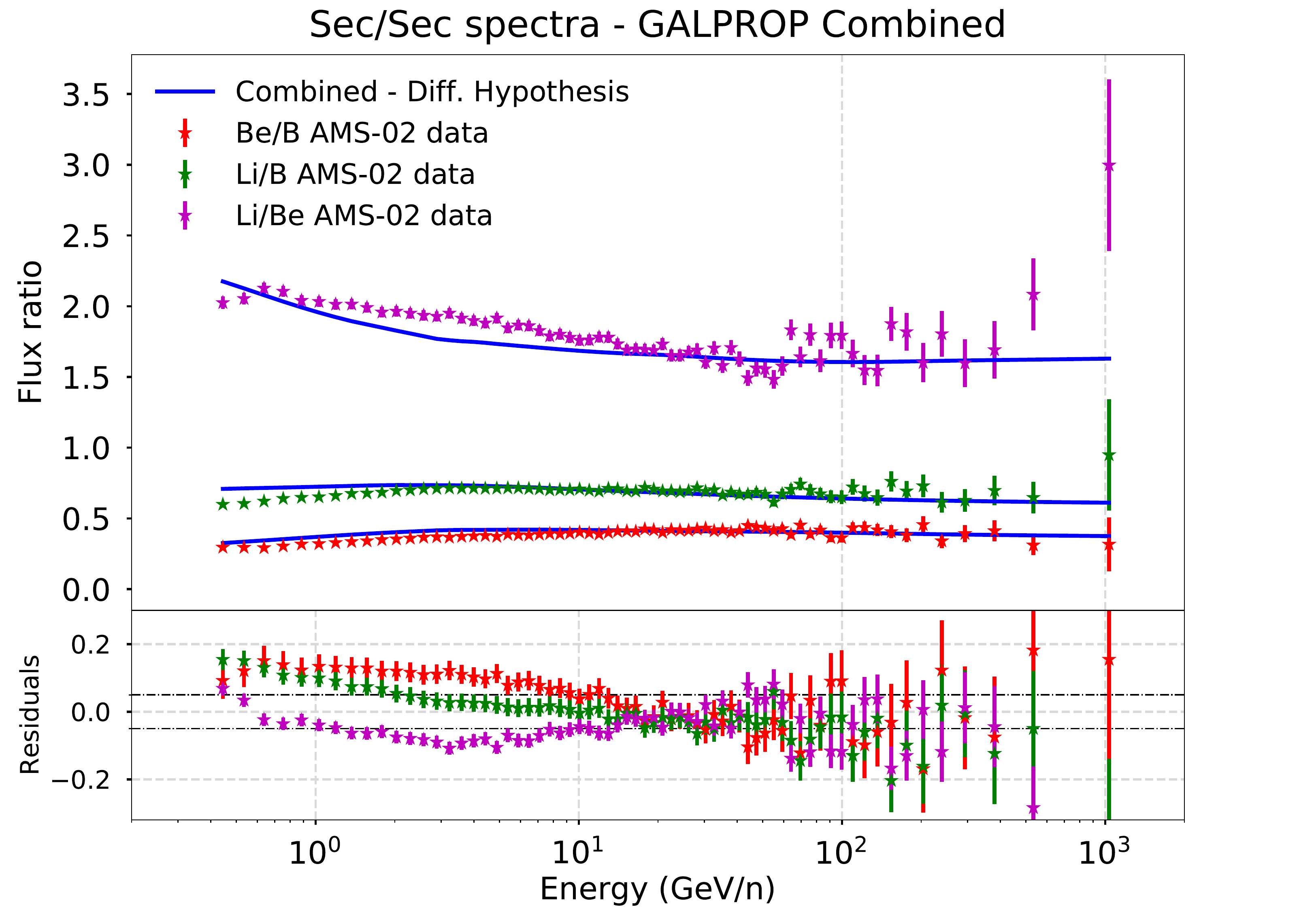}
\rightskip -0.3cm
\caption{Secondary-over-secondary flux ratios of Li, Be and B computed using the propagation parameters and scaling factors found in the combined analysis for each of the cross sections sets compared to the AMS-02 experimental data. The grid lines at the level of $5\%$ residuals are highlighted in the lower panel in a different color in order to guide the eye.}
	\label{fig:SecSec_combined}
\end{figure}

In Figure~\ref{fig:SecSec_combined} we show a comparison of the predicted secondary-to-secondary spectra of Li, Be and B with the AMS-02 data using the propagation parameters and scaling factors obtained in the combined analysis for each of the cross sections sets.
As we see, the secondary-over-secondary ratios predicted with the {\tt DRAGON2} cross sections are consistent with data within the level of $5\%$ (highlighted residual grid lines). The Li/Be and Be/B ratios are the ones with slightly larger discrepancies with respect to experimental data
. In the case of the {\tt GALPROP} parametrizations, the discrepancies at low energy are larger (often above $5\%$ at low energy) and the energy dependence of the predicted Be/B ratio seems to be slightly different to that shown by the experimental AMS-02 Be/B ratio. We remind that these discrepancies could change when evaluating the secondary-to-secondary flux ratios using a different gas distribution.

In order to investigate how important this bias is in the determination of the propagation parameters, we explore, in the next section, a different procedure that could allow us to obtain a more reliable set of propagation parameters
after correcting their production cross sections.

\subsection{``Scaled'' analysis}
\label{sec:CXSanal}

To prevent the propagation parameters from being biased, we have implemented an alternative strategy. This procedure allows us to use the propagation parameters extracted from the B ratios, since the amount of data for the cross sections of B production allows a better parametrization of these channels.

\begin{table}[t]
\centering
\begin{tabular}{|c|c|c|c|c|c|c|c|}

\cline{2-7} \multicolumn{1}{c|}{ } & \multicolumn{2}{|c|}{\textbf{\large{$\mathcal{S}_{B}$}} }& \multicolumn{2}{|c|}{\textbf{\large{$\mathcal{S}_{Be}$}} } & \multicolumn{2}{|c|}{\textbf{\large{$\mathcal{S}_{Li}$}} } \\ 
\cline{2-7}
\multicolumn{1}{c|} { }& \textbf{Fit} & \textbf{Comb} & \textbf{Fit} & \textbf{Comb} & \textbf{Fit} & \textbf{Comb} \\ \hline
{\tt DRAGON2} & 1.04 & 1.05  &  0.975  & 0.985  & 0.945  &  0.97 \\ \hline 
{\tt GALPROP} & 0.94 & 1.01  &  0.865  & 0.895 & 1.18  &  1.26  \\ \hline 
\end{tabular}
\caption{Comparison between the scaling factors inferred in the combined analyses (Comb) and in the fit of the secondary-over-secondary ratios measured by AMS-02 (Fit) for the cross sections sets used. These scaling factors are the same for both diffusion hypotheses} used. The $1\sigma$ statistical errors on this determination are $\sim 1\%$.
\label{tab:Scaling_comp}
\end{table}

First, we fit the AMS-02 secondary-over-secondary ratios by adjusting the scaling factors of the B, Be and Li production cross sections, including also the nuisance terms (eq. 2.8), and using the diffusion parameters obtained in the B/O analysis of each cross sections set (see Figure~\ref{fig:boxplot_Standard}). We use the diffusion parameters inferred from this ratio as reference since we expect that the prediction of the flux ratios of B to be more accurate than for the other secondary CRs. In fact B is mainly produced from interactions of C and O projectiles, and the interactions channels of B production are those with more cross sections measurements available. In addition, the fraction of secondary oxygen is negligible, so that the oxygen nuclei entering in this ratio are pure primaries in the full energy range (see ref.~\cite{CarmeloBlasi}). 
Then, we apply the same independent MCMC analysis to each of the ratios and compare the propagation parameters.The idea behind the first step is that all possible scalings that fit the AMS-02 secondary-over-secondary ratios and lay inside cross sections data uncertainties are allowed candidates, but the preferred one is that combination of scaling factors which involves the minimum change from the original parametrization. In order to carry out this analysis we use again a MCMC procedure, in which the free parameters are the scaling factors, $\mathcal{S}_B$, $\mathcal{S}_{Be}$ and $\mathcal{S}_{Li}$, and the data to be fitted are just the secondary-over-secondary ratios above $20 \units{GeV/n}$, with a penalty factor proportional to the nuisance term of eq.~\ref{eq:myNuisance}.
This will avoid biasing the scaling factors obtained by the cross sections parametrizations when combining the different species. 

In Table~\ref{tab:Scaling_comp}, a comparison among the scaling factors obtained with both analysis approaches is shown.
As we see, while the scaling factors from both procedures are quite similar for the {\tt DRAGON2} cross sections, they are considerably different for the {\tt GALPROP} cross sections. This indicates that the bias due to the shape of the cross sections can be relevant. In fact, these differences are higher than the $\pm$4\% uncertainties related to variations in the diffusion coefficient that we assume, although they are lower than $\pm$8\%. 
We stress the fact that scaling factors for the {\tt DRAGON2} cross sections are <$5.5\%$ and that, as expected, the B scaling is the closest to 1 in the {\tt GALPROP} parametrization.

\begin{figure}[!t]
\centering
\includegraphics[width=0.51\textwidth, height=0.245\textheight]{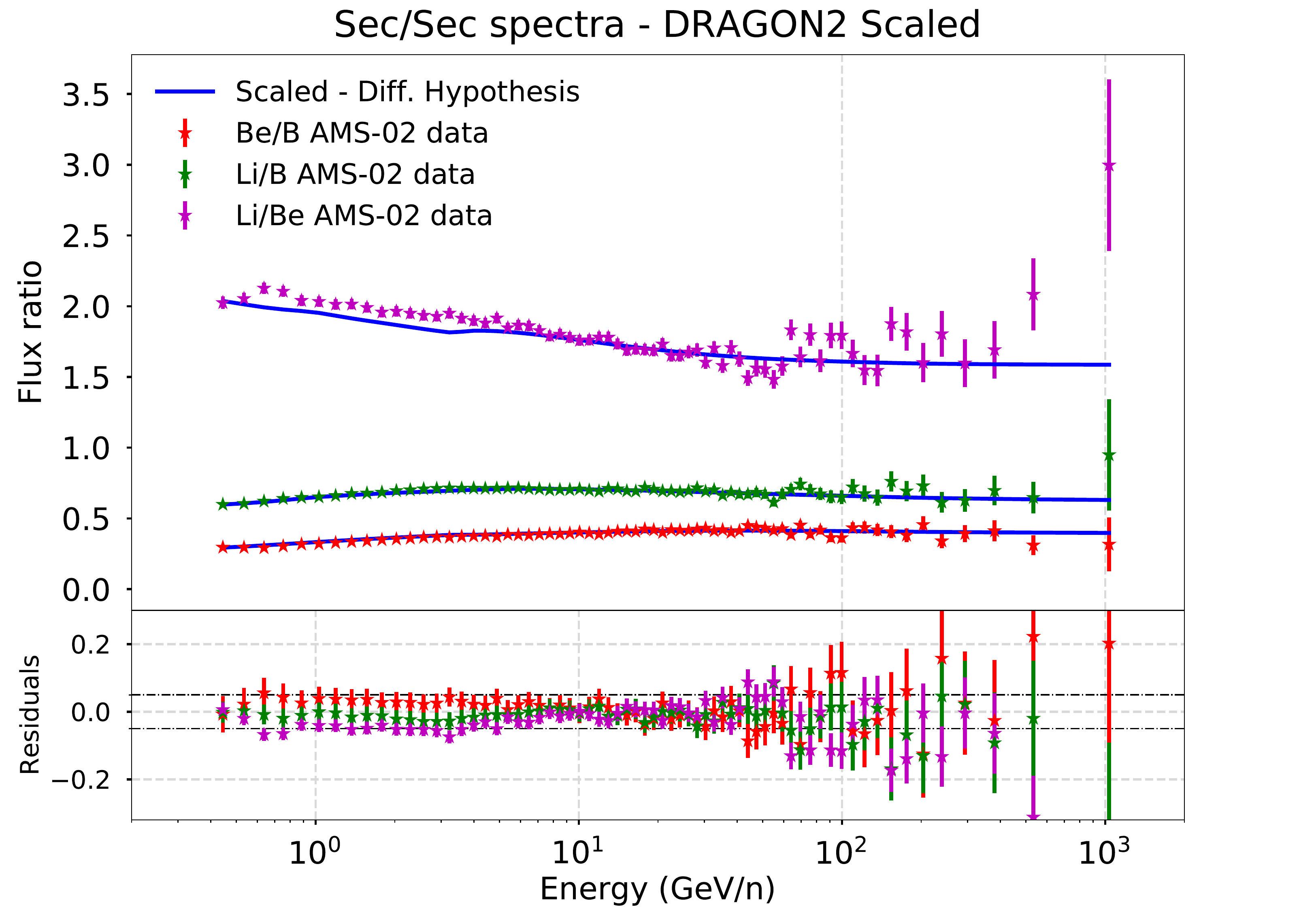} \hspace{-0.3cm}
\includegraphics[width=0.51\textwidth, height=0.245\textheight]{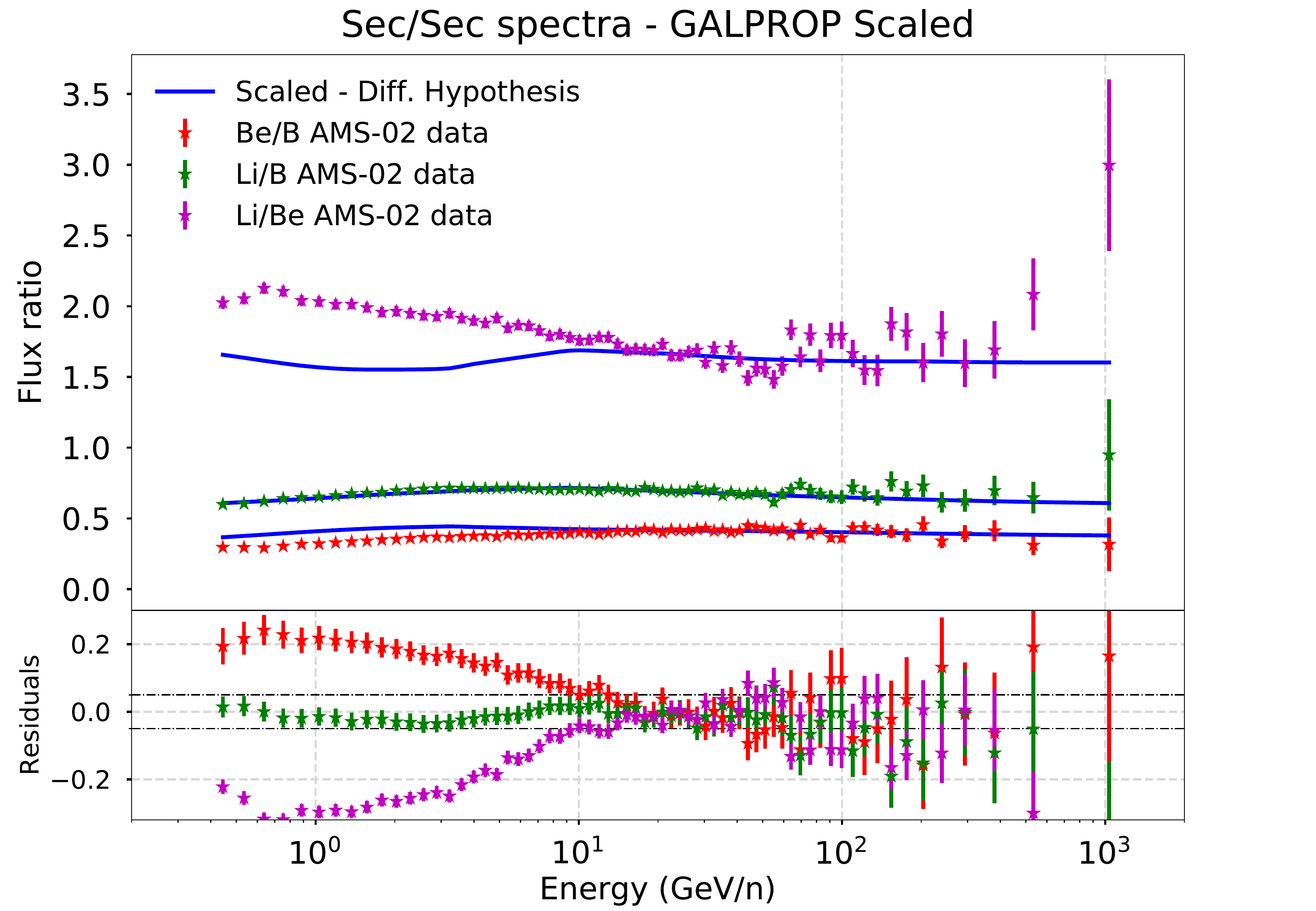}
\rightskip -0.3cm
\caption{Secondary-over-secondary flux ratios predicted by the scaled {\tt DRAGON2} and {\tt GALPROP} cross sections parametrizations compared to the AMS-02 data. They are fitted to reproduce the high energy part of the spectra, while the low energy part of the Li/Be and Be/B ratios mainly depends on the halo size value used.}
	\label{fig:SecSec_DR}
\end{figure}

\begin{figure}[!tb]
\centering
\includegraphics[width=\textwidth, height=0.59\textheight]{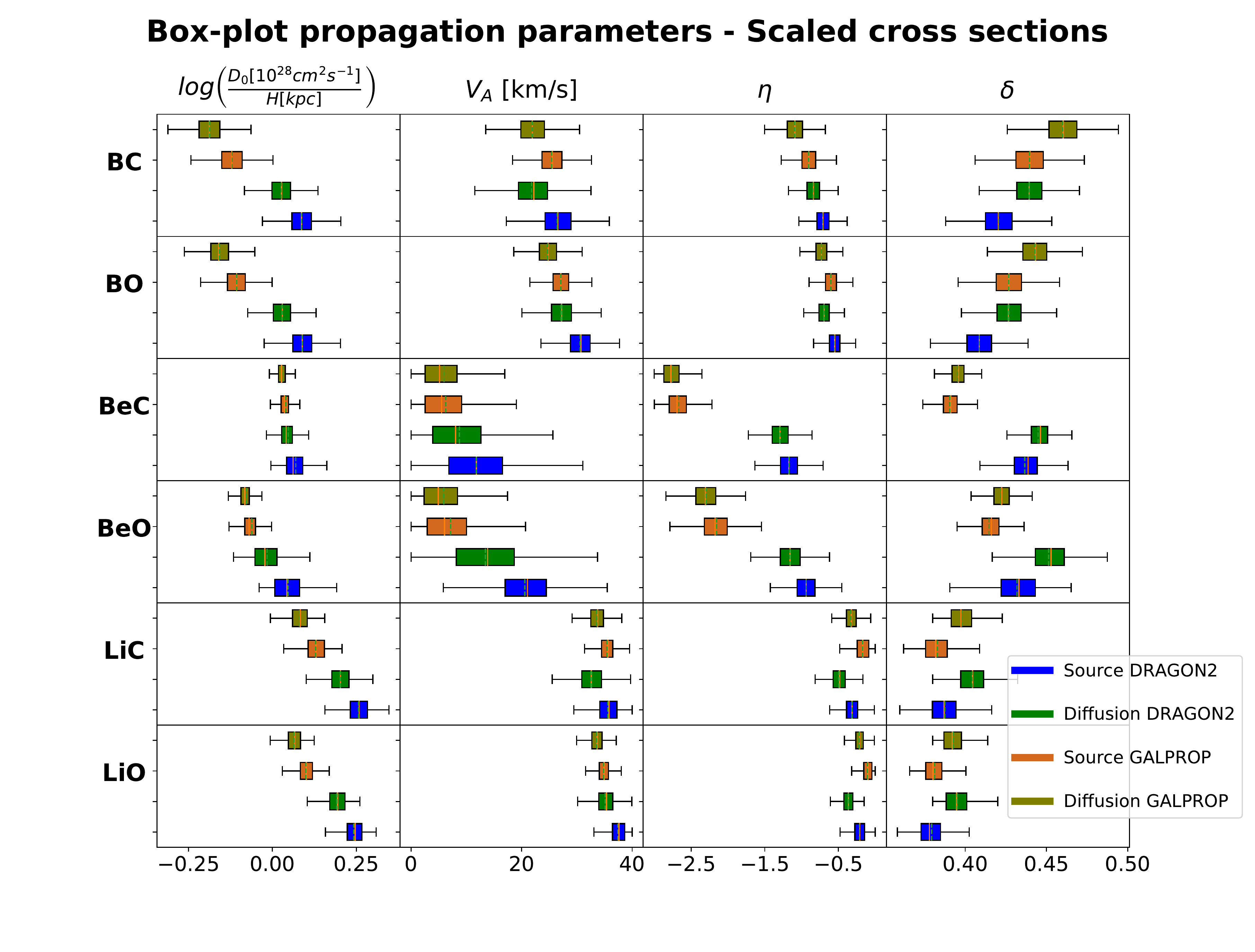}
\caption{Box-plots representation of the diffusion parameters obtained in the independent analyses of the Li,Be,B/C,O AMS-02 ratios for the scaled {\tt GALPROP} and {\tt DRAGON2} cross sections for both hypothesis on the diffusion coefficient. The meaning of the lines and of the boxes is the same as in Fig.~\ref{fig:boxplot_Standard}.}
\label{fig:boxplot_Renorm}
\end{figure}

These scaled secondary-over-secondary flux ratios (figure~\ref{fig:SecSec_DR}) are expected to show larger discrepancies at low energies, while the high energy part should be well reproduced. 
Then, with these scaling factors, the independent analysis of each of the ratios is performed for the {\tt DRAGON2} and {\tt GALPROP} cross sections and the results are shown in Figure~\ref{fig:boxplot_Renorm}. Again, the results are summarized in the tables~\ref{tab:PDF_single_scaled_Source} and~\ref{tab:PDF_scaled_Diff} in the appendix~\ref{sec:appendixA}.

It is worth remarking here that analyzing the ratios after scaling the cross sections does not lead to just a change in the predicted normalization of the diffusion coefficient ($D_0$), but also to slight changes in the slope ($\delta$) and in the low-energy parameters (although never by more than 2$\sigma$).
As we can see, the {\tt GALPROP} cross sections predict different propagation parameters for each secondary  CR species, meaning that the analysis must be performed including more degrees of freedom for the shape of their cross sections energy spectrum. This is probably related to the fact that the Be and Li scaling factors deviate more than $10\%$, since the cross sections at low energy are always better constrained than at high energy. As expected, the diffusion parameters inferred from the ratios of B are the most reliable. On the other hand, we see that for the {\tt DRAGON2} parametrizations, both the $D_0$ and $\delta$ values are nearly identical for the B and Be ratios, while they are significantly different for the Li ratios. 
From these results, we can see that a $\delta$ value of $0.43-0.45$ is favored when excluding the Li predictions, and this value goes down to an average of $\sim$0.41 when including it.

\section{Conclusions}
\label{MCMC_conc}

In this work, we have studied the two parametrizations of the diffusion coefficient that best-fit with the theoretical expectations and are suitable to reproduce the experimental data for the two most accurate cross sections parametrizations  (namely {\tt DRAGON2} and {\tt GALPROP} cross sections). We have presented the MCMC analyses of the secondary-over-primary and secondary-over-secondary flux ratios, for the Li, Be and B, in order to determine the propagation parameters that best reproduce AMS-02 data for the two cross sections sets and both diffusion hypotheses (with and without high energy break). This combination of different secondary CRs provides a way to mitigate the effect of cross sections uncertainties in the determination of the propagation parameters, which are mostly dominated by their normalization in the cross sections parametrizations.

In general, we see that the propagation parameters obtained in the independent analysis are very similar for the cross sections studied, favoring a $\delta \sim 0.37 - 0.45$, specially hard for Li ratios.
The combined analysis tends to $\delta \sim 0.41-0.44$ for the {\tt DRAGON2} cross sections, while $\delta=0.37-0.40$ for the {\tt GALPROP} cross sections. In addition, we have demonstrated that the secondary-over-primary ratios of B, Be and Li to C and O can be simultaneously reproduced with low discrepancies including a renormalization of the production cross sections. We underline that the {\tt DRAGON2} cross sections yield an almost perfect fit of all these ratios with respect to the AMS-02 and Voyager-1 data at the same time with very low corrections on the normalization of the B, Be and Li production cross sections. On the other hand,  the {\tt GALPROP} cross sections lead to large discrepancies that are due to the shape of the cross sections energy spectra.
Nevertheless, the propagation parameters determined in this combined analysis could be biased by deviations in the adjustment of the energy dependence of the cross sections parametrizations from the different secondary CRs, as it may be the case of the combined analysis with the {\tt GALPROP} parametrizations, specially favoring the values found from the Li ratios. 

In order to investigate and prevent this bias, which might be particularly important for parametrizations based on experimental cross sections data, because of the uncertainty on the cross sections of Li production, an alternative procedure has been explored to determine the diffusion parameters differently. After applying the analysis presented in section~\ref{sec:CXSanal}, for the {\tt GALPROP} and {\tt DRAGON2} cross sections, we have determined that the shape of the energy spectrum for the cross sections of Li production needs to be slightly adjusted in order to get the same prediction for the diffusion parameters in all the ratios. For the {\tt DRAGON2} spallation cross sections, the inferred diffusion parameters are compatible within $1\sigma$ for the diffusion hypothesis of the diffusion coefficient for the Be and B with $\delta \sim 0.43-0.45$. In turn, when using the {\tt GALPROP} cross sections, we infer more disperse values of $\delta \sim 0.37-0.46$. It seems that a change on the shape of the {\tt GALPROP} cross sections of Be and Li production is needed, which is probably related to the fact that both the Li and Be cross sections energy spectra need scaling factors deviating of more than $10\%$ to reproduce the secondary-over-secondary ratios. When excluding the predictions from the Li ratios, both the {\tt DRAGON2} and {\tt GALPROP} cross sections favor a larger $\delta$ value, of around $0.44$ for the {\tt DRAGON2} parametrizations and $\sim0.42$ for the {\tt GALPROP} ones. Isotopic cosmic-ray data will be a valuable tool to adjust the different spallation cross section parametrizations in the various isotopic channels.


\acknowledgments
We remark the crucial help of Daniele Gaggero in the development of the paper and his examinations during the evolution of the ideas commented here.

Many thanks to the instituto de física teórica (IFT) in Madrid for hosting Pedro De la Torre for a long stay there and specially to the DAMASCO (DArk MAtter AStroparticles and COsmology) group for their support and valuable conversations related to this work.

This work has been carried out using the RECAS computing infrastructure in Bari (\url{https://www.recas-bari.it/index.php/en/}). A particular acknowledgment goes to G. Donvito and A. Italiano for their valuable support.

\bibliographystyle{apsrev4-1}
\bibliography{biblio}

\newpage
\appendix

\section{Summary of the MCMC results}
\label{sec:appendixA}

\begin{figure}[!htpb]
	\centering
	\includegraphics[width=0.78\textwidth, height=0.47\textheight]{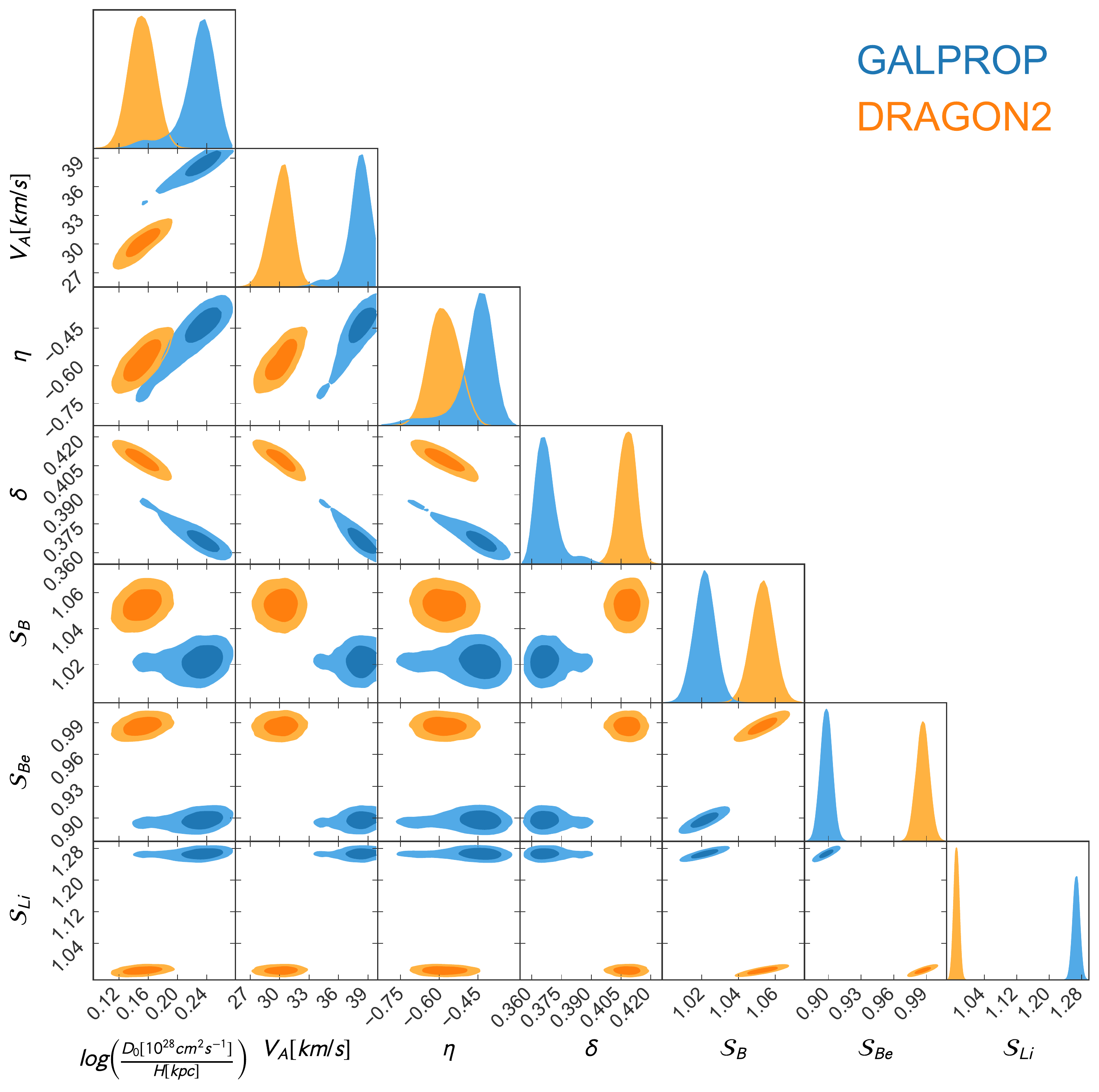}
	\caption{Corner plot representing the PDF of every parameter included in the combined MCMC analysis with the diffusion coefficient parametrization of eq. \ref{eq:sourcehyp} for the {\tt DRAGON2} and {\tt GALPROP} cross sections parametrizations. The inner panels represent the correlations between parameters with the 1 and 2 $\sigma$ surface probabilities highlighted.}
	\label{fig:Corner_Source}
\end{figure}

\begin{figure}[!htpb]
	\centering
	\includegraphics[width=0.8\textwidth, height=0.47\textheight]{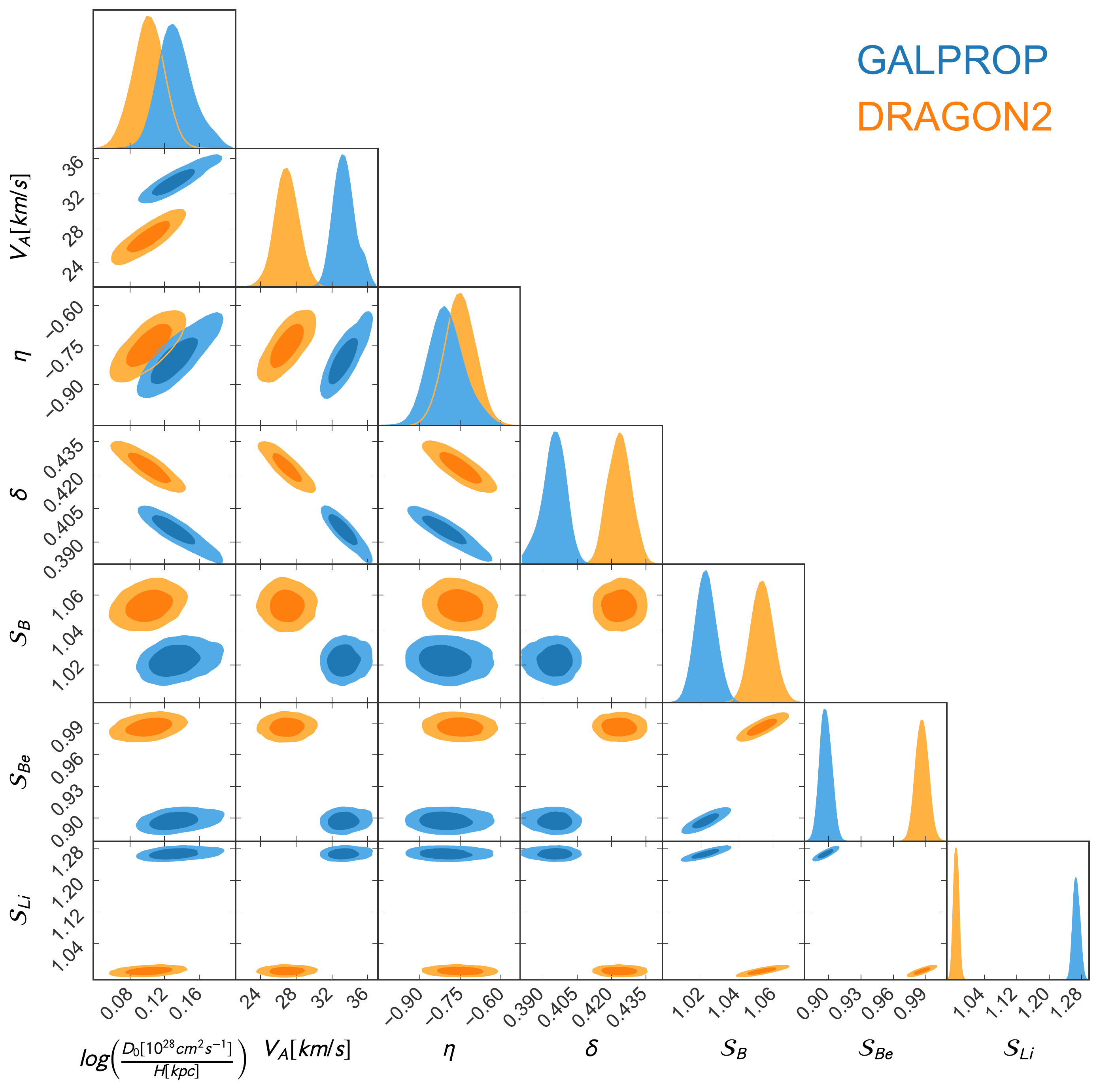}
	\caption{Corner plot as in figure \ref{fig:Corner_Source} but for the diffusion coefficient parametrization of eq.~\ref{eq:breakhyp}.}
	\label{fig:Corner_Diff}
\end{figure}

In this appendix, the corner plots obtained from the MCMC algorithm of the combined analysis for each cross sections set and both diffusion parametrizations used ('Source' and 'Diffusion' hypotheses) are shown (figs.~\ref{fig:Corner_Source} and~\ref{fig:Corner_Diff}) to visualize the probability distribution functions (PDFs) of each propagation parameter and the correlations between them. These corner plots contain the diffusion parameters ($D_0$, $\delta$, $\eta$ and $V_A$) as well as the scaling factors included as nuisance parameters.   Furthermore, we are summarizing the results for the PDFs of each of the propagation parameters found in form of tables. First, we show the results for the combined analysis, then, the parameters found in the independent analyses of each secondary-to-primary ratio (B/C, B/O, Be/C, Be/O, Li/C, Li/O) and, finally, the analysis of each of the ratios after scaling the original cross sections according to the secondary-over-secondary ratios. The tables contain the median value (coinciding with the maximum posterior probability value) $\pm$ the $1\sigma$ uncertainty related to its determination, assuming their PDFS to be Gaussian, and the actual range of values contained in the 95\% probability region. 

\vspace{1cm}
\begin{center}
    \textbf{\large{Propagation parameters PDF - Source hypothesis}}
\end{center} 

\newpage 

\begin{center}
    \textbf{\large{Propagation parameters PDF - Diffusion hypothesis}}
\end{center} 

\vspace{1.1cm}
\begin{table}[htb!]
\centering
\textbf{\Large{Combined analysis}}

\vspace{0.4cm}

\resizebox*{0.72\columnwidth}{0.17\textheight}{
\begin{tabular}{|c|c|c|c|c|c|}
\hline
 \multicolumn{1}{|c|} { } & \multicolumn{2}{c|}{\textbf{\large{Diffusion hypothesis}}} & \multicolumn{2}{c|}{\textbf{\large{Source hypothesis}}}  \\
\cline{2-5} &{\tt DRAGON2} & {\tt GALPROP} & {\tt DRAGON2} & {\tt GALPROP} \\ 
\hline
\multirow{2}{7em}{\centering $\frac{D_0 \, (10^{28}\units{cm^{2}/s})}{H\, (kpc)} $} & 1.11 $\pm$ 0.02 & 1.14 $\pm$ 0.03   &  1.16 $\pm$ 0.02 & 1.26  $\pm$ 0.03 \\  &  [1.06, 1.30]  &   [1.08, 1.30] &  [1.11, 1.15] &   [1.21, 1.32]\\ \hline 

\multirow{2}{6em}{\centering\large{$V_A$ (km/s)}}  & 26.92  $\pm$ 1.28 & 33.32 $\pm$ 1.23 &  30.20 $\pm$ 1.28 &  38.21 $\pm$ 1.11 \\  &  [24.52, 29.48]  &   [31.03, 35.78] & [27.80, 32.17] &   [35.99, 40.01] \\ \hline 

\multirow{2}{6em}{\centering\large{$\eta$}}    & -0.75 $\pm$ 0.06 & -0.80 $\pm$ 0.07 &  -0.58 $\pm$ 0.06 & -0.44 $\pm$ 0.7 \\ &  [-0.87, -0.65]  &   [-0.92, -0.66] &  [-0.68, -0.46] &   [-0.58, -0.36]\\ \hline 

\multirow{2}{6em}{\centering\large{$\delta$}}  & 0.42 $\pm$ 0.01 & 0.40 $\pm$ 0.01 &  0.41 $\pm$ 0.01 & 0.37 $\pm$ 0.01  \\  &  [0.39, 0.44]  &   [0.38, 0.42] &  [0.38, 0.43] &   [0.35, 0.39] \\ \hline 
\end{tabular}
}
\caption{Results obtained for the combined ratios analysis for each cross sections sets and diffusion coefficient parametrizations. Maximum posterior probability value with its standard deviation (assuming the PDF to be a Gaussian) and the range of parameters within $95\%$ of confidence level are shown for each parameter.}
\label{tab:PDF_Combined}
\end{table}

\begin{table}[htb!]
\centering
\textbf{\Large{Independent analyses - Source hypothesis}}

\vspace{0.4cm}

\resizebox*{0.81\columnwidth}{0.17\textheight}{
\begin{tabular}{|c|c|c|c|c|c|c|c|}
\hline
\multicolumn{7}{|c|}{\textbf{\large{{\tt DRAGON2} parameters - Source hypothesis}}} \\
\hline &\textbf{B/C} & \textbf{B/O} & \textbf{Be/C} & \textbf{Be/O} & \textbf{Li/C} & \textbf{Li/O} \\ 
\hline
\multirow{2}{7em}{\centering $\frac{D_0 \, (10^{28}\units{cm^{2}/s})}{H\, (kpc)}$} & 1.02 $\pm$ 0.05 & 1.04$\pm$ 0.05   &  1.09 $\pm$ 0.05 &  1.06 $\pm$ 0.07 & 1.37  $\pm$ 0.05 &  1.33 $\pm$ 0.04  \\  &  [0.94., 1.12]  &   [0.95, 1.14] &   [1.03, 1.19]  &  [0.96, 1.19] &   [1.27, 1.45] &   [1.25, 1.39] \\ \hline 

\multirow{2}{6em}{\centering\large{$V_A$ (km/s)}}  & 25.72  $\pm$ 3.22 & 20.65 $\pm$ 2.65 &  11.74 $\pm$ 7.24  &  29.65 $\pm$ 7.19 &  36.49 $\pm$ 2.68 &  38.12 $\pm$ 1.66 \\  &  [19.72, 32.16.]  &   [24.85, 34.95] &   [0., 26.26]  &  [15.27, 40.75] &   [31.07, 40.31] &   [34.9, 40.66] \\ \hline 

\multirow{2}{6em}{\centering\large{$\eta$}}    & -0.66 $\pm$ 0.12 & -0.51 $\pm$ 0.11 &  -1.15 $\pm$ 0.16 &  -0.94 $\pm$ 0.19 & -0.37 $\pm$ 0.11 &  -0.29 $\pm$ 0.08  \\ &  [-0.80, -0.42]  &   [-0.72, -0.33] &   [-1.47, -0.83]  &  [-1.32, -0.6] &   [-0.59, -0.15] &   [-0.45, -0.15] \\ \hline 

\multirow{2}{6em}{\centering\large{$\delta$}}  & 0.43 $\pm$ 0.01 & 0.41 $\pm$ 0.01 &  0.44 $\pm$ 0.01  &  0.43 $\pm$ 0.02 & 0.39 $\pm$ 0.01 &  0.38 $\pm$ 0.01  \\  &  [0.41, 0.45]  &   [0.39, 0.43] &   [0.41, 0.46]  &  [0.39, 0.47] &   [0.37, 0.41] &   [0.36, 0.40] \\ \hline 
\end{tabular}
}

\vspace{0.6 cm}
\resizebox*{0.81\columnwidth}{0.17\textheight}{
\begin{tabular}{|c|c|c|c|c|c|c|c|}
\hline
\multicolumn{7}{|c|}{\textbf{\large{{\tt GALPROP} parameters - Source hypothesis}}} \\
\hline &\textbf{B/C} & \textbf{B/O} & \textbf{Be/C} & \textbf{Be/O} & \textbf{Li/C} & \textbf{Li/O} \\ 
\hline
\multirow{2}{7em}{\centering $\frac{D_0 \, (10^{28}\units{cm^{2}/s})}{H\, (kpc)} $} & 0.99 $\pm$ 0.04 & 0.98 $\pm$ 0.04   &  1.26 $\pm$ 0.03 &  1.10 $\pm$ 0.05 & 0.98 $\pm$ 0.05 &  1.00 $\pm$ 0.05  \\  &  [0.92, 1.07]  &   [0.92, 1.07] &   [1.22, 1.32]  &  [1.05, 1.15] &   [0.88, 1.08] &   [0.91, 1.07] \\ \hline 

\multirow{2}{6em}{\centering\large{$V_A$ (km/s)}}  & 27.17  $\pm$ 2.61 & 29.11 $\pm$ 2.14 &  6.31 $\pm$ 7.20  &  10.50 $\pm$ 7.48 &  32.37 $\pm$ 2.04 &  32.40 $\pm$ 1.66 \\  &  [21.95, 32.25]  &   [25.07, 33.39] &   [0., 15.05]  &  [0., 25.46] &   [28.65, 36.45] &   [29.28, 35.04] \\ \hline 

\multirow{2}{6em}{\centering\large{$\eta$}}    & -0.54 $\pm$ 0.12 & -0.29 $\pm$ 0.09 &  -1.56 $\pm$ 0.19 &  -1.15 $\pm$ 0.19 & -0.6 $\pm$ 0.20 &  -0.32 $\pm$ 0.14  \\ &  [-0.78, -0.30]  &   [-0.45, -0.11] &   [-1.94, -1.18]  &  [-1.53, -0.78] &   [-1.0, -0.44] &   [-0.60, -0.10] \\ \hline 
\multirow{2}{6em}{\centering\large{$\delta$}}  & 0.43 $\pm$ 0.01 & 0.42 $\pm$ 0.01 &  0.39 $\pm$ 0.01  &  0.42 $\pm$ 0.01 & 0.38 $\pm$ 0.01 &  0.37 $\pm$ 0.01  \\  &  [0.41, 0.45]  &   [0.40, 0.45] &   [0.37, 0.41]  &  [0.40, 0.44] &   [0.36, 0.40] &   [0.36, 0.39] \\ \hline 
\end{tabular}
}

\caption{Results obtained in the independent analysis of the flux ratios B/C, Be/C, Li/C, B/O, Be/O and Li/O for the diffusion coefficient parametrization of eq. \ref{eq:sourcehyp} for each of the cross sections sets analysed.}
\label{tab:PDF_single_Source}
\end{table}

\vspace{0.2cm}

\begin{table}[htb!]
\centering
\vspace{0.4cm}
\textbf{\Large{Independent analyses - Diffusion hypothesis}}

\vspace{0.4cm}

\resizebox*{0.81\columnwidth}{0.18\textheight}{
\begin{tabular}{|c|c|c|c|c|c|c|c|}
\hline
\multicolumn{7}{|c|}{\textbf{\large{ {\tt DRAGON2} parameters - Diffusion hypothesis}}} \\
\hline &\textbf{B/C} & \textbf{B/O} & \textbf{Be/C} & \textbf{Be/O} & \textbf{Li/C} & \textbf{Li/O} \\ 
\hline
\multirow{2}{7em}{\centering $\frac{D_0 \, (10^{28}\units{cm^{2}/s})}{H\, (kpc)} $} & 0.97 $\pm$ 0.04 & 0.98$\pm$ 0.04   &  1.07 $\pm$ 0.03 &  1.00 $\pm$ 0.05 & 1.30  $\pm$ 0.04 &  1.28 $\pm$ 0.05  \\  &  [0.90., 1.05]  &   [0.91, 1.06] &   [1.03, 1.14]  &  [0.92, 1.11] &   [1.21, 1.37] &   [1.18, 1.36] \\ \hline 

\multirow{2}{6em}{\centering\large{$V_A$ (km/s)}}  & 22.02  $\pm$ 3.54 & 26.53 $\pm$ 2.54 &  7.60 $\pm$ 6.70  &  14.19 $\pm$ 8.76 &  33.31 $\pm$ 2.72 &  36.60 $\pm$ 2.47 \\  &  [14.94, 28.86.]  &   [21.45, 31.43] &   [0., 20.00]  &  [0, 29.27] &   [27.87, 37.49] &   [31.64, 40.00] \\ \hline 

\multirow{2}{6em}{\centering\large{$\eta$}}    & -0.78 $\pm$ 0.13 & -0.65 $\pm$ 0.10 &  -1.30 $\pm$ 0.16 &  -1.16 $\pm$ 0.21 & -0.54 $\pm$ 0.12 &  -0.38 $\pm$ 0.11  \\ &  [-1.04, -0.54]  &   [-0.75, -0.55] &   [-1.60, -0.98]  &  [-1.58, -0.74] &   [-0.78, -0.36] &   [-0.60, -0.22] \\ \hline 

\multirow{2}{6em}{\centering\large{$\delta$}}  & 0.44 $\pm$ 0.01 & 0.43 $\pm$ 0.01 &  0.45 $\pm$ 0.01  &  0.45 $\pm$ 0.02 & 0.41 $\pm$ 0.01 &  0.39 $\pm$ 0.01  \\  &  [0.42, 0.46]  &   [0.41, 0.45] &   [0.43, 0.47]  &  [0.41, 0.47] &   [0.39, 0.43] &   [0.37, 0.41] \\ \hline 
\end{tabular}
}

\vspace{0.6 cm}
\resizebox*{0.81\columnwidth}{0.18\textheight}{
\begin{tabular}{|c|c|c|c|c|c|c|c|}
\hline
\multicolumn{7}{|c|}{\textbf{\large{{\tt GALPROP} parameters - Diffusion hypothesis}}} \\
\hline &\textbf{B/C} & \textbf{B/O} & \textbf{Be/C} & \textbf{Be/O} & \textbf{Li/C} & \textbf{Li/O} \\ 
\hline
\multirow{2}{7em}{\centering $\frac{D_0 \, (10^{28}\units{cm^{2}/s})}{H\, (kpc)} $} & 0.94 $\pm$ 0.04 & 0.94 $\pm$ 0.03   &  1.24 $\pm$ 0.02 &  1.08 $\pm$ 0.03 & 0.89 $\pm$ 0.04 &  0.91 $\pm$ 0.03  \\  &  [0.87, 1.04]  &   [0.88, 1.00] &   [1.21, 1.28]  &  [1.04, 1.15] &   [0.81, 0.97] &   [0.85, 0.96] \\ \hline 

\multirow{2}{6em}{\centering\large{$V_A$ (km/s)}}  & 24.44  $\pm$ 2.97 & 27.08 $\pm$ 1.91 &  4.74 $\pm$ 4.99  &  7.64 $\pm$ 7.23 &  29.11 $\pm$ 2.20 &  29.32 $\pm$ 1.52 \\  &  [18.90, 29.54]  &   [23.26, 30.86] &   [0., 14.72]  &  [0., 22.10] &   [24.71, 33.21] &   [26.28, 31.76] \\ \hline 

\multirow{2}{6em}{\centering\large{$\eta$}}    & -0.66 $\pm$ 0.12 & -0.39 $\pm$ 0.09 &  -1.68 $\pm$ 0.16 &  -1.27 $\pm$ 0.19 & -0.94 $\pm$ 0.20 &  -0.61 $\pm$ 0.11  \\ &  [-0.90, -0.18]  &   [-0.57, -0.23] &   [-1.98, -1.36]  &  [-1.65, -0.91] &   [-1.34, -0.62] &   [-0.83, -0.41] \\ \hline 
\multirow{2}{6em}{\centering\large{$\delta$}}  & 0.45 $\pm$ 0.01 & 0.44 $\pm$ 0.01 &  0.40 $\pm$ 0.01  &  0.42 $\pm$ 0.01 & 0.40 $\pm$ 0.01 &  0.39 $\pm$ 0.01  \\  &  [0.43, 0.47]  &   [0.42, 0.46] &   [0.38, 0.43]  &  [0.40, 0.44] &   [0.37, 0.42] &   [0.37, 0.41] \\ \hline 
\end{tabular}
}

\caption{Same as in \ref{tab:PDF_single_Source} but for the diffusion coefficient parametrization of eq. \ref{eq:breakhyp}.}
\label{tab:PDF_single_Diff}
\end{table}

\begin{table}
\centering

\textbf{\Large{Independent scaled analyses - Source hypothesis}}

\vspace{0.4cm}

\resizebox*{0.81\columnwidth}{0.17\textheight}{
\begin{tabular}{|c|c|c|c|c|c|c|c|}
\hline
\multicolumn{7}{|c|}{\textbf{\large{{\tt DRAGON2} parameters - Source hypothesis}}} \\
\hline &\textbf{B/C} & \textbf{B/O} & \textbf{Be/C} & \textbf{Be/O} & \textbf{Li/C} & \textbf{Li/O} \\ 
\hline
\multirow{2}{7em}{\centering $\frac{D_0 \, (10^{28}\units{cm^{2}/s})}{H\, (kpc)} $} & 1.09 $\pm$ 0.05 & 1.09$\pm$ 0.05   &  1.06 $\pm$ 0.02 &  1.04 $\pm$ 0.06 & 1.30  $\pm$ 0.05 &  1.28 $\pm$ 0.05  \\  &  [1.0, 1.19]  &   [1.01, 1.19] &   [1.01, 1.16]  &  [0.94, 1.16] &   [1.20, 1.40] &   [1.18, 1.35] \\ \hline 

\multirow{2}{6em}{\centering\large{$V_A$ (km/s)}}  & 26.58  $\pm$ 3.54 & 30.72 $\pm$ 2.78 &  11.80 $\pm$ 7.05  &  21.06 $\pm$ 6.06 &  35.81 $\pm$ 2.55 &  37.63 $\pm$ 1.82 \\  &  [18.5, 33.16]  &   [25.16, 35.58] &   [0., 25.48]  &  [8.96, 30.15] &   [30.71, 40.01] &   [32.99, 40.50] \\ \hline 

\multirow{2}{6em}{\centering\large{$\eta$}}    & -0.71 $\pm$ 0.12 & -0.55 $\pm$ 0.11 &  -1.17 $\pm$ 0.18 &  -0.94 $\pm$ 0.18 & -0.32 $\pm$ 0.11 &  -0.21 $\pm$ 0.09  \\ &  [-0.95, -0.57]  &   [-0.77, -0.35] &   [-1.24, -0.81]  &  [-1.28, -0.58] &   [-0.54, -0.10] &   [-0.41, -0.03] \\ \hline 

\multirow{2}{6em}{\centering\large{$\delta$}}  & 0.42 $\pm$ 0.01 & 0.41 $\pm$ 0.01 &  0.44 $\pm$ 0.01  &  0.43 $\pm$ 0.02 & 0.39 $\pm$ 0.01 &  0.38 $\pm$ 0.01  \\  &  [0.40, 0.44]  &   [0.39, 0.43] &   [0.41, 0.45]  &  [0.39, 0.46] &   [0.37, 0.41] &   [0.36, 0.41] \\ \hline 
\end{tabular}
}

\vspace{0.6 cm}
\resizebox*{0.81\columnwidth}{0.17\textheight}{
\begin{tabular}{|c|c|c|c|c|c|c|c|}
\hline
\multicolumn{7}{|c|}{\textbf{\large{{\tt GALPROP} parameters - Source hypothesis}}} \\
\hline &\textbf{B/C} & \textbf{B/O} & \textbf{Be/C} & \textbf{Be/O} & \textbf{Li/C} & \textbf{Li/O} \\ 
\hline
\multirow{2}{7em}{\centering $\frac{D_0 \, (10^{28}\units{cm^{2}/s})}{H\, (kpc)} $} & 0.89 $\pm$ 0.04 & 0.93$\pm$ 0.04   &  0.90 $\pm$ 0.02 &  0.93 $\pm$ 0.03 & 1.14  $\pm$ 0.04 &  1.11 $\pm$ 0.03  \\  &  [0.81, 0.98]  &   [0.86, 0.94] &   [1.01, 1.08]  &  [0.87, 1.01] &   [1.06, 1.22] &   [1.05, 1.17] \\ \hline 

\multirow{2}{6em}{\centering\large{$V_A$ (km/s)}}  & 25.56  $\pm$ 2.80 & 27.09 $\pm$ 2.16 &  5.50 $\pm$ 5.33  &  6.06 $\pm$ 6.28 &  35.61 $\pm$ 1.71 &  34.94 $\pm$ 1.30 \\  &  [19.96, 30.70.]  &   [22.93, 31.41] &   [0., 16.16]  &  [0, 18.72] &   [32.19, 37.17] &   [32.34, 37.0] \\ \hline 

\multirow{2}{6em}{\centering\large{$\eta$}}    & -0.90 $\pm$ 0.14 & -0.60 $\pm$ 0.11 &  -2.69 $\pm$ 0.20 &  -2.17 $\pm$ 0.25 & -0.17 $\pm$ 0.11 &  -0.10 $\pm$ 0.08  \\ &  [-1.17, -0.62]  &   [-0.82, -0.38] &   [-2.99, -2.29]  &  [-2.58, -1.67] &   [-0.39, +0.05] &   [-0.26, +0.04] \\ \hline 

\multirow{2}{6em}{\centering\large{$\delta$}}  & 0.44 $\pm$ 0.01 & 0.43 $\pm$ 0.01 &  0.39 $\pm$ 0.01  &  0.42 $\pm$ 0.02 & 0.38 $\pm$ 0.01 &  0.38 $\pm$ 0.01  \\  &  [0.41, 0.46]  &   [0.41, 0.45] &   [0.37, 0.41]  &  [0.40, 0.44] &   [0.36, 0.41] &   [0.36, 0.41] \\ \hline 
\end{tabular}
}

\caption{Results obtained in the independent analysis of the flux ratios B/C, Be/C, Li/C, B/O, Be/O and Li/O for the diffusion coefficient parametrization of eq. \ref{eq:sourcehyp} with the scaled cross sections sets analysed. }
\label{tab:PDF_single_scaled_Source}
\end{table}

\vspace{1.7cm}

\begin{table}[htb!]
\centering
\textbf{\Large{Independent scaled analyses - Diffusion hypothesis}}

\vspace{0.4cm}

\resizebox*{0.81\columnwidth}{0.18\textheight}{
\begin{tabular}{|c|c|c|c|c|c|c|c|}
\hline
\multicolumn{7}{|c|}{\textbf{\large{{\tt DRAGON2} parameters - Diffusion hypothesis}}} \\
\hline &\textbf{B/C} & \textbf{B/O} & \textbf{Be/C} & \textbf{Be/O} & \textbf{Li/C} & \textbf{Li/O} \\ 
\hline
\multirow{2}{7em}{\centering $\frac{D_0 \, (10^{28}\units{cm^{2}/s})}{H\, (kpc)} $} & 1.03 $\pm$ 0.04 & 1.03$\pm$ 0.04   &  1.04 $\pm$ 0.03 &  0.98 $\pm$ 0.05 & 1.22  $\pm$ 0.04 &  1.22 $\pm$ 0.04  \\  &  [0.95, 1.11]  &   [0.95, 1.11]  &   [1.0, 1.11]  &  [0.92, 1.09] &   [1.14, 1.30] &   [1.14, 1.30] \\ \hline 

\multirow{2}{6em}{\centering\large{$V_A$ (km/s)}}  & 22.20  $\pm$ 4.29 & 27.25 $\pm$ 2.76 &  8.60 $\pm$ 6.77  &  13.84 $\pm$ 8.22 &  32.61 $\pm$ 2.66 &  35.36 $\pm$ 2.12 \\  &  [13.62, 27.32]  &   [21.73, 32.47] &   [0., 22.14]  &  [0, 29.27] &   [27.29, 37.83] &   [31.08, 38.56] \\ \hline 

\multirow{2}{6em}{\centering\large{$\eta$}}    & -0.84 $\pm$ 0.13 & -0.69 $\pm$ 0.10 &  -1.29 $\pm$ 0.16 &  -1.16 $\pm$ 0.20 & -0.48 $\pm$ 0.13 &  -0.37 $\pm$ 0.09  \\ &  [-1.10, -0.6]  &   [-0.89, -0.49] &   [-1.61, -0.97]  &  [-1.36, -0.78] &   [-0.74, -0.26] &   [-0.75, -0.19] \\ \hline 

\multirow{2}{6em}{\centering\large{$\delta$}}  & 0.44 $\pm$ 0.01 & 0.43 $\pm$ 0.01 &  0.45 $\pm$ 0.01  &  0.45 $\pm$ 0.02 & 0.40 $\pm$ 0.01 &  0.39 $\pm$ 0.01  \\  &  [0.42, 0.46]  &   [0.41, 0.45] &   [0.43, 0.47]  &  [0.42, 0.47] &   [0.39, 0.42] &   [0.36, 0.41] \\ \hline 
\end{tabular}
}

\vspace{0.6 cm}
\resizebox*{0.81\columnwidth}{0.18\textheight}{
\begin{tabular}{|c|c|c|c|c|c|c|c|}
\hline
\multicolumn{7}{|c|}{\textbf{\large{{\tt GALPROP} parameters - Diffusion hypothesis}}} \\
\hline &\textbf{B/C} & \textbf{B/O} & \textbf{Be/C} & \textbf{Be/O} & \textbf{Li/C} & \textbf{Li/O} \\ 
\hline
\multirow{2}{7em}{\centering $\frac{D_0 \, (10^{28}\units{cm^{2}/s})}{H\, (kpc)} $} & 0.83 $\pm$ 0.04 & 0.85 $\pm$ 0.04   &  1.02 $\pm$ 0.03 &  0.92 $\pm$ 0.03 & 1.08 $\pm$ 0.03 &  1.07 $\pm$ 0.03  \\  &  [0.78, 0.92]  &   [0.79, 0.92] &   [0.98, 1.07]  &  [0.88, 0.96] &   [1.02, 1.14] &   [1.01, 1.12] \\ \hline 

\multirow{2}{6em}{\centering\large{$V_A$ (km/s)}}  & 21.95  $\pm$ 3.14 & 24.82 $\pm$ 2.31 &  5.14 $\pm$ 4.83  &  4.91 $\pm$ 5.46 &  33.77 $\pm$ 1.85 &  33.67 $\pm$ 1.39 \\  &  [15.75, 28.23]  &   [19.20, 29.04] &   [0., 14.80]  &  [0., 15.83] &   [30.07, 36.73] &   [30.89, 36.15] \\ \hline 

\multirow{2}{6em}{\centering\large{$\eta$}}    & -1.09 $\pm$ 0.16 & -0.73 $\pm$ 0.11 &  -2.70 $\pm$ 0.17 &  -2.31 $\pm$ 0.21 & -0.32 $\pm$ 0.10 &  -0.21 $\pm$ 0.07  \\ &  [-1.44, -0.79]  &   [-0.57, -0.23] &   [-2.96, -2.36]  &  [-2.69, -1.90] &   [-0.52, -0.12] &   [-0.35, -0.07] \\ \hline 

\multirow{2}{6em}{\centering\large{$\delta$}}  & 0.46 $\pm$ 0.01 & 0.44 $\pm$ 0.01 &  0.40 $\pm$ 0.01  &  0.42 $\pm$ 0.01 & 0.40 $\pm$ 0.01 &  0.39 $\pm$ 0.01  \\  &  [0.44, 0.48]  &   [0.42, 0.47] &   [0.38, 0.41]  &  [0.40, 0.43] &   [0.38, 0.42] &   [0.37, 0.40] \\ \hline 
\end{tabular}
}

\caption{Same as in \ref{tab:PDF_single_scaled_Source} but for the diffusion coefficient parametrization of eq. \ref{eq:breakhyp}.}
\label{tab:PDF_scaled_Diff}
\end{table}

\FloatBarrier
Finally, here we report tables with the $\chi^2$ values obtained with the best fits parameters of the independent and combined analyses for the {\tt GALPROP} and {\tt DRAGON2} cross sections. As previously reported, the inclusion of a break in the injection improves the quality for all the simulated ratios. By comparing the $\chi^2$ values for both hypotheses in the independent analysis, we observe that both cross sections yield similarly good fits. Interestingly, the maximum $\Delta \chi^2$ between both cross sections is observed for the Be ratios. This could be due to the halo size value used for each parametrization but it could also be related to the cross sections of Be production by themselves. 

Besides, we observe that the difference of $\chi^2$ values between the source and diffusion hypotheses in the independent analyses is of $\Delta \chi^2 \sim 10$ for the ratios of Li and B (roughly unaffected by the halo size value), both for {\tt DRAGON2} and {\tt GALPROP}. Applying Wilk's theorem we can translate this difference in $\chi^2$ values to significance, finding it to be of around $4.7 \sigma$.

Finally, from these tables we show that the combined analyses yield better $\chi^2$ values for the {\tt DRAGON2} cross sections than for the {\tt GALPROP} ones as displayed in~\ref{fig:SecPrim_DRAGON2} and~\ref{fig:SecPrim_GALPROP}.

\hskip 0.4cm

\begin{table}[htb!]
\centering
\resizebox*{0.73\columnwidth}{0.17\textheight}{
\begin{tabular}{|c|c|c|c|c|c|c|c|}
\hline
\multicolumn{7}{|c|}{\textbf{\large{$\chi^2$ values of best-fit parameters $-$ GALPROP}}} \\
\hline &\textbf{B/C} & \textbf{B/O} & \textbf{Be/C} & \textbf{Be/O} & \textbf{Li/C} & \textbf{Li/O} \\ 
\hline
\multirow{2}{10em}{\centering\large{Source hypothesis Indep. analysis}} & \multirow{2}{*}{41.79} & \multirow{2}{*}{37.22}  &  \multirow{2}{*}{44.94} &  \multirow{2}{*}{43.6} & \multirow{2}{*}{40.27} &  \multirow{2}{*}{36.82}  \\ & & & & & & \\  \hline 

\multirow{2}{10em}{\centering\large{Diffusion hypothesis Indep. analysis}}  & \multirow{2}{*}{31.87} & \multirow{2}{*}{26.42}  &  \multirow{2}{*}{47.08} &  \multirow{2}{*}{43.03} & \multirow{2}{*}{32.42} &  \multirow{2}{*}{27.64}  \\ & & & & & & \\  \hline 

\multirow{2}{10em}{\centering\large{Source hypothesis Comb. analysis}}  & \multirow{2}{*}{92.79} & \multirow{2}{*}{152.35}  &  \multirow{2}{*}{176.64} &  \multirow{2}{*}{141.12} & \multirow{2}{*}{69.78} &  \multirow{2}{*}{42.58}  \\ & & & & & & \\  \hline 

\multirow{2}{10em}{\centering\large{Diffusion hypothesis Comb. analysis}} & \multirow{2}{*}{86.87} & \multirow{2}{*}{154.06}  &  \multirow{2}{*}{169.79} &  \multirow{2}{*}{138.94} & \multirow{2}{*}{64.56} &  \multirow{2}{*}{34.86}  \\ & & & & & & \\  \hline 
\end{tabular}}
\caption{$\chi^2$ values of the best-fit diffusion parameters found for the MCMC analyses with the {\tt GALPROP} cross sections.}
\label{tab:Xi2_GALPROP}
\end{table}

\vspace{0.6 cm}
\begin{table}[htb!]
\centering
\resizebox*{0.73\columnwidth}{0.17\textheight}{
\begin{tabular}{|c|c|c|c|c|c|c|c|}
\hline
\multicolumn{7}{|c|}{\textbf{\large{$\chi^2$ values of best-fit parameters $-$ DRAGON2}}} \\
\hline &\textbf{B/C} & \textbf{B/O} & \textbf{Be/C} & \textbf{Be/O} & \textbf{Li/C} & \textbf{Li/O} \\ 
\hline
\multirow{2}{10em}{\centering\large{Source hypothesis Indep. analysis}} & \multirow{2}{*}{40.8} & \multirow{2}{*}{37.48}  &  \multirow{2}{*}{35.61} &  \multirow{2}{*}{33.25} & \multirow{2}{*}{39.76} &  \multirow{2}{*}{39.96}  \\ & & & & & & \\  \hline 

\multirow{2}{10em}{\centering\large{Diffusion hypothesis Indep. analysis}}  & \multirow{2}{*}{31.55} & \multirow{2}{*}{27.3}  &  \multirow{2}{*}{33.94} &  \multirow{2}{*}{31.98} & \multirow{2}{*}{30.37} &  \multirow{2}{*}{29.04}  \\ & & & & & & \\  \hline 

\multirow{2}{10em}{\centering\large{Source hypothesis Comb. analysis}}  & \multirow{2}{*}{41.69} & \multirow{2}{*}{38.47}  &  \multirow{2}{*}{55.42} &  \multirow{2}{*}{53.68} & \multirow{2}{*}{44.14} &  \multirow{2}{*}{52.46}  \\ & & & & & & \\  \hline 

\multirow{2}{10em}{\centering\large{Diffusion hypothesis Comb. analysis}} & \multirow{2}{*}{32.43} & \multirow{2}{*}{29.04}  &  \multirow{2}{*}{54.01} &  \multirow{2}{*}{51.51} & \multirow{2}{*}{34.24} &  \multirow{2}{*}{41.69}  \\ & & & & & & \\  \hline 
\end{tabular}}
\caption{Same as in table~\ref{tab:Xi2_GALPROP}, but for the {\tt DRAGON2}.}
\label{tab:Xi2_DRAGON2}
\end{table}

\vspace{0.6 cm}

\end{document}